\documentclass[aps,prd,preprintnumbers,showpacs,superscriptaddress,nofootinbib,amsmath,amssymb,floats,floatfix,showkeys,notitlepage,longbibliography]{revtex4-1}
\usepackage{comment}
\usepackage{graphicx}
\usepackage{subfigure}
\usepackage{palatino}
\usepackage[commandnameprefix=always]{changes}
\usepackage{hyperref}
\hypersetup{colorlinks=true,linkcolor=red,urlcolor=red,citecolor=red}
\usepackage[toc,page]{appendix}
\usepackage[normalem]{ulem}

\usepackage{lipsum}
\usepackage{graphicx}
\usepackage{subfigure}
\usepackage{palatino}
\usepackage{float}
\usepackage{sans}
\usepackage{adjustbox}
\usepackage{latexsym}
\usepackage{amsmath}
\usepackage{amssymb}
\usepackage{amsfonts}
\usepackage{dcolumn}
\usepackage{bm}
\usepackage{tikz}
\usepackage{bigints}
\usepackage{array,tabularx,multirow,booktabs}
\usepackage[tracking=true]{microtype}
\SetTracking{}{500}
\SetTracking{encoding={*}, shape=sc}{40}
\allowdisplaybreaks
\usepackage{adjustbox}
\usepackage{latexsym}
\usepackage{amsmath}
\usepackage{amssymb}
\usepackage{amsfonts}
\usepackage{dcolumn}
\usepackage{bm}
\usepackage{tikz}
\usepackage{bigints}
\usepackage{array,tabularx,multirow,booktabs}
\usepackage[tracking=true]{microtype}
\usepackage{color}
\allowdisplaybreaks

\begin{document}

\title{A Deep Dive into classical and Topological CFT Thermodynamics in Lifshitz and Hyperscaling Violating Black Holes}

\author{Mohammad Ali S. Afshar}
\email{m.a.s.afshar@gmail.com}
\affiliation{Department of Physics, Faculty of Basic
Sciences, University of Mazandaran\\ P. O. Box 47416-95447, Babolsar, Iran}
\affiliation{School of Physics, Damghan University, P. O. Box 3671641167, Damghan, Iran}

\affiliation{Canadian Quantum Research Center, 204-3002 32 Ave Vernon, BC V1T 2L7, Canada}

\author{Mohammad Reza Alipour}
\email{mr.alipour@stu.umz.ac.ir}
\affiliation{School of Physics, Damghan University, P. O. Box 3671641167, Damghan, Iran}
\affiliation{Department of Physics, Faculty of Basic
Sciences, University of Mazandaran\\ P. O. Box 47416-95447, Babolsar, Iran}

\author{Saeed Noori Gashti}
\email{saeed.noorigashti@stu.umz.ac.ir; saeed.noorigashti70@gmail.com}
\affiliation{School of Physics, Damghan University, P. O. Box 3671641167, Damghan, Iran}

\author{Jafar Sadeghi}
\email{pouriya@ipm.ir}
\affiliation{Department of Physics, Faculty of Basic
Sciences, University of Mazandaran\\ P. O. Box 47416-95447, Babolsar, Iran}
\affiliation{Canadian Quantum Research Center, 204-3002 32 Ave Vernon, BC V1T 2L7, Canada}

\vspace{1.5cm}\begin{abstract}
To effectively utilize the AdS/CFT correspondence, a precise set of rules must be established to guide the translation of computed quantities in the gravitational sector into their CFT counterparts, and vice versa. This framework is commonly referred to as the holographic dictionary. The formulation of such dictionaries opens a two-way gateway, allowing researchers to extend theoretical principles and findings from one domain into the other for further exploration and study. The development of a holographic dictionary for Lifshitz black holes and hyperscaling violation (HSV) models \cite{6} has provided an essential foundation for studying CFT thermodynamics and phase behavior of these black holes. Based on this framework, we will investigate their thermodynamic properties using two distinct approaches. In the first step, we adopt the classical and traditional method, identifying critical points to examine the behavior of the free energy function as a function of temperature near the critical boundary. By analyzing its behavior, we will study phase transitions and then proceed to evaluate the stability of the models. In the next step, to compare both methodologies and highlight their equivalence—particularly demonstrating the accessibility of the topological method compared to the classical approach—we will analyze phase behavior through the lens of topological charges.
\end{abstract}

\date{\today}



\maketitle
\tableofcontents
\newpage
\section{Introduction}
The gauge/gravity duality, or AdS/CFT correspondence, states that a gravitational theory formulated in an asymptotically Anti-de Sitter (AdS) space is equivalent to a conformal field theory (CFT) defined at its boundary. This means that every phenomenon occurring in the gravitational sector within the AdS framework—including black hole physics—can be described equivalently within the dual CFT.  This correspondence serves as a powerful tool, akin to a golden key, for studying complex quantum dynamics in strongly coupled systems, such that calculations that are difficult to perform directly in gravitational theory may be more manageable in the corresponding CFT, and vice versa. The existence of this duality allows researchers to strategically tackle challenging problems in high-energy physics and quantum gravity with greater flexibility. 
Despite its intricate nature, the core concept of this correspondence seems to stem from a fundamental logical principle: any phenomena occurring in higher-dimensional spaces must remain interpretable within lower dimensions at the boundary, subject to constraints. To aids in intuition and conceptual understanding and a simplified analogy, one could compare this idea to classical mathematical principles such as Gauss's divergence theorem (which translates three-dimensional to two-dimensional interpretations) and Stokes’ theorem (mapping two-dimensional to one-dimensional properties), both of which find applications in electromagnetic gauge fields (Gauss's and Ampère’s laws).\\  
However, beyond the theoretical foundation, the critical aspect lies in the practical realization and application of this duality. To make use of this correspondence, a precise set of rules must be established to guide the translation of quantities computed in the gravity sector to their CFT counterparts, and vice versa. These translation rules are commonly referred to as the "holographic dictionary". For example, the partition function of the gravitational theory can be equated to the generating function of the dual CFT \cite{1}.  
Among the most compelling applications of the AdS/CFT framework is its usage in the study of black hole physics, which are considered ideal scenarios for coupling gravity to gauge fields, such as electromagnetism. Despite the experimental inaccessibility and the predominantly theoretical nature of black hole studies, they remain remarkable candidates for investigating this duality. In particular, black hole thermodynamics, given its minimal reliance on relativistic complexities, may serve as the optimal context. On one hand, when the CFT is at a finite temperature, it represents a thermal system—analogous to standard thermodynamics in any field theory. On the other hand, in the AdS context, black holes exhibit well-defined thermodynamic properties such as:\\
$\bullet$ Temperature, derived from Hawking radiation\\ 
$\bullet$ Entropy, associated with the event horizon area\\
$\bullet$ Energy (or mass), computed using general relativity and quantum field theory in curved spacetime.\\
Thus, leveraging our golden duality, we can effectively translate the thermodynamic behavior of an AdS black hole into the language of its dual CFT. Under this framework:\\
$\bullet$ The black hole temperature in the bulk corresponds to the CFT temperature in finite volume.\\
$\bullet$ The entropy computed from the black hole’s horizon maps onto the thermal entropy in the CFT. \\
Also, Similar correspondences can be established for other thermodynamic variables.  For instance:\\
$\bullet$ The gravitational action computed on the boundary of black hole spacetime can be linked to the free energy of the thermal dual CFT.\\
$\bullet$ In holographic CFTs dual to Einstein gravity, the central charge is related to both the AdS radius and Newton’s constant, implying that changes in central charge within the CFT can lead to variations in both the cosmological constant and Newton’s constant in the gravity sector. Consequently, varying the cosmological constant on the gravity side is dual to varying the central charge or the number of colors in the dual gauge theory. Therefore, the central charge and its conjugate chemical potential (also known as color susceptibility) play an essential role in the holographic dual of extended black hole thermodynamics \cite{2,2',2''}. However, not everything follows this direct mapping, as certain complications arise. For example, bulk pressure is not dual to the pressure of the field theory, or the Black hole thermodynamic volume does not correspond to the spatial volume of the CFT formulation \cite{3}.  Fortunately, these issues can be addressed by introducing new thermodynamic variables, which help refine the CFT interpretation of extended black hole thermodynamics \cite{4,4'}. While the derivation of these holographic dictionaries is inherently complex, they unlock new perspectives for studying lower-dimensional boundary physics.  For instance, in the Gauss–Bonnet model, the formulation of a holographic dictionary enabled investigations into CFT thermodynamics across different dimensions, yielding valuable insights into dimensional constraints \cite{5,5'}.  
In this regard, recent studies aimed at constructing a holographic dictionary for Lifshitz and hyperscaling violating black holes \cite{6} motivated us to investigate the CFT thermodynamics of these black holes from two perspectives:\quad
1. The general and traditional method, 2. A novel topological approach. Most black hole models proposed for study are inherently theoretical due to the lack of direct experimental verification. Based on prior knowledge, experience, and available data, researchers hope that if spacetime geometry follows Riemannian principles and the prevailing mechanics remain relativistic, certain constructed models will closely align with cosmic reality. Given these assumptions, it is natural to expect such models to exhibit familiar physical behaviors. However, what makes theoretical black hole models particularly intriguing is the potential emergence of novel gravitational-gauge behaviors, which arise from the interplay between gravitational corrections and modern gauge fields. These behaviors may not form in classical, Earth-bound conditions but can manifest under extreme astrophysical scenarios. And that's what motivates us to study these behaviors. Black hole thermodynamics is no exception to this phenomenon. For instance, it has been widely observed that AdS black holes, where the cosmological constant is treated as pressure \cite{2}, exhibit distinct small-to-large black hole phase transitions, akin to the gas-liquid transition in Van der Waals fluids \cite{7,8,9}. Additionally, the generalized Euler relation is recovered in restricted phase space \cite{10}.\\  
Accordingly, in the general and traditional method, parameters in the CFT space are first defined according to the holographic dictionary, followed by the computation of critical points and the Gibbs free Energy, enabling the analysis of phase behavior based on temperature.\\
Meanwhile, a recently developed approach introduces a mapping of the energy function into a two-dimensional space $( r, \Theta )$, opening a new avenue for studying black hole behavior. This technique allows for the investigation of vector field behavior in mapped space concerning the computation of winding numbers, yielding interesting insights into geodesic structures and black hole thermodynamics. The study of black hole thermodynamics, particularly the analysis of phase transitions, has been approached from multiple perspectives to deepen our understanding of these complex systems. Among these, topological thermodynamics has emerged as a powerful and insightful framework. This approach leverages topological concepts to characterize the global properties and classification of black hole phases, going beyond conventional methods that focus primarily on local thermodynamic quantities. By examining the topological structure of thermodynamic parameter spaces, researchers are able to identify critical points, classify phase transitions, and gain a more unified understanding of black hole behavior across different gravitational backgrounds. This perspective has proven especially valuable in exploring rich phenomena such as first- and second-order phase transitions, critical phenomena, and universality classes in a wide range of black hole solutions.. Building on Duan’s topological current theory, these approaches uncover global structures within the thermodynamic phase space. Notably, two complementary techniques—the temperature-based (T) method and the free energy-based (F) method—have been developed by S.W. Wei and collaborators. The T method identifies critical points via topological charges by treating temperature as a vector field. In contrast, the F method investigates the Helmholtz free energy landscape, using curvature and winding properties to distinguish between first-and second-order transitions. Stability analysis within the F framework relies on the second derivative of free energy. Together, these methods provide a unified framework for analyzing the rich phase structure of black holes. For more study, see \cite{a19,a20,20a,21a,22a,23a,24a,25a,26a,27a,28a,29a,31a,33a,34a,35a,37a,38',38a,38b,38c,39a,40a,41a,42a,43a,44a,44c,44d,44e,44f,44g,44h,44i,44j,44k,44l,44m,44n,44o,44p,45p,46p,47p,48p,49p,50p,51p,52p,53p,54p,55p,56p}.\\
To illustrate the behavioral similarities and provide a visual comparison of both approaches—and recognizing that in cases where one method (traditional or topological) may be difficult to apply, the other could serve as a suitable alternative— we will employ both approaches in analyzing CFT thermodynamics for Lifshitz and hyperscaling-violating black holes. We will also investigate black holes across various parametric scenarios, assessing potential differences in behavior, analyzing their phase dynamics, and ultimately proposing a behavioral classification based on topological charges.\\
Accordingly, in this article, we will begin by introducing the models, their general parameters, and constraints. We will then proceed with the traditional thermodynamic approach, analyzing phase diagrams based on swallowtail structures or smooth continuous phase transitions. Subsequently, we will explore the stability and topological method and its application in black hole thermodynamics, and finally, we will synthesize and conclude our findings. 
\section{Lifshitz and hyperscaling violating black holes}
 Lifshitz black holes emerged in 2008–2009 as gravitational solutions modeling anisotropic scaling between time and space, inspired by Lifshitz fixed points in condensed matter physics. In these early models, the bulk geometry was altered to display an anisotropic scaling between time and space—represented by a dynamical critical exponent "z", where time scales as $t\to\lambda^z t$ and space as $x\to\lambda x$. This idea was initially driven by the need to construct gravitational duals for certain condensed matter systems, where such anisotropic scaling is a defining feature. Kachru and his colleagues, in the foundational work, introduced these spacetimes in string theory to holographically describe non-relativistic quantum critical systems \cite{51,52}. 
In early 2010, researchers recognized that many strongly correlated systems showed even more complex scaling behavior that could not be captured by the exponent "z" alone. This led to the introduction of an additional parameter
—the hyperscaling violation exponent(HSV) \(\theta\) to capture broader scaling violations in thermodynamic properties, often arising in Einstein-Maxwell-Dilaton theories studied by researchers like Dong, Harrison, and Charmousis \cite{53}. Including \(\theta\) in the gravitational models modifies the overall scaling of the metric so that physical quantities in the dual field theory scale anomalously. These hyperscaling-violating black holes allow one to incorporate richer thermodynamic and hydrodynamic behaviors, making them especially relevant for studying phase transitions, transport phenomena, and quantum criticality in non-relativistic systems.
These two solutions extend the AdS/CFT correspondence to systems without conformal symmetry, such as strange metals or high-temperature superconductors. Lifshitz metrics model anisotropic criticality, while hyperscaling violation modifies thermodynamic scaling, crucial for describing real-world condensed matter systems. They enable studies of transport properties (e.g., conductivity), entanglement entropy, and phase transitions in holographic frameworks.\\
An interesting point to note is that they can also coexist ($z\neq1,\theta\neq0$), but the distinction lies in whether the emphasis is on anisotropic spacetime symmetry (Lifshitz) or effective dimensionality (HSV).
\begin{center}
\textbf{Lifshitz black holes  features}
\end{center}
For the spherical form of this black hole, Table I presents the parametric constraints along with their corresponding physical implications.
\begin{center}
\begin{table}[H]
  \centering
\begin{tabular}{|p{1cm}|p{4cm}|p{5cm}|p{2.5cm}|p{4cm}|}
  \hline
  \centering{Case}  & \centering{Physical Meaning} &\centering{Thermo condition}& \centering{Holographic Dual}& \hspace{1.2cm} Stability\\[3mm]
   \hline
  \centering{$z=1$} & \centering {Relativistic AdS black holes} & \centering{Specific heat can be positive (stable)} & \centering{CFTs} & Stable(AdS/CFT framework)\\[3mm]
   \hline
 \centering{$z>1$} & \centering {Spacetime becomes non-relativistic (time scales faster than space)} & \centering{Specific heat can be positive (stable) if spatial dimensions d compensate} &\centering {Lifshitz field theories}&\ Stable if NEC holds \\[3mm]
   \hline
   \centering{$z<1$} & \centering {Time scales slower than space, leading to potential instabilities, Unphysical or exotic} & \centering{Specific heat often becomes negative, leading to thermodynamic instability} & \centering {Rarely viable} &\ Unstable (ghosts, tachyons) \\[3mm]
   \hline
   \end{tabular}
   \caption{Lifshitz black holes }\label{1}
\end{table}
 \end{center}
\begin{center}
\textbf{HSV black holes  features}
\end{center}
Furthermore, Table II presents the parametric constraints along with their corresponding physical implications with respect to the spherical form of a black hole.
\begin{center}
\begin{table}[H]
  \centering
\begin{tabular}{|p{2cm}|p{3cm}|p{3cm}|p{3.5cm}|p{4cm}|}
  \hline
  \centering{Case}  & \centering{Physical Meaning} &\centering{Thermo condition}& \centering{Holographic Dual}& \hspace{1.2cm}Stability\\[3mm]
   \hline
  \centering{$z=1$,$\theta=0$} & \centering {Relativistic AdS black holes} & \centering{Specific heat can be positive (stable)} & \centering{CFTs} & Stable(AdS/CFT framework)\\[3mm]
   \hline
 \centering{$z>1$,$\theta=0$} & \centering {Spacetime becomes non-relativistic (time scales faster than space)} & \centering{Specific heat can be positive (stable) if spatial dimensions d compensate} &\centering {Lifshitz field theories}&\ Stable if NEC holds \\[3mm]
   \hline
   \centering{$0<\theta<d$,\\$z\geq1+\theta/d$} & \centering {Hyperscaling violation,Mimics quantum critical systems with effective dimensional reduction } & \centering{NEC condition:\\$(d-\theta)(z-1-\theta/d)\geq0$} , \quad $(z - 1)(d + z - \theta) \geq$ 0 & \centering {Dual to quantum critical systems violating hyperscaling.eg: spin liquids, disordered phases.} &\ depends to condition \\[3mm]
   \hline
   \centering{$\theta=d$ \\z Not allowed} & \centering {Naked singularity or Unphysical spacetime} & \centering{NEC condition:Violated} & \centering {No consistent dual theory} &\ Unphysical  \\[3mm]
   \hline
   \centering{$\theta<0$} & \centering {Requires exotic matter (e.g., nonlinear gauge fields)} & \centering{Depends on the chosen value of z (e.g. always holds for $z>1$ and $d>\theta$)} & \centering {Speculative duals: Hypothetical "emergent" higher-dimensional theories (e.g., glassy phases).} &\ Often unstable. \\[3mm]
   \hline  
   \end{tabular}
   \caption{HSV black holes}\label{1}
\end{table}
 \end{center}

The general line element describing Lifshitz and hyperscaling violating black holes in \((d + 1)\)-dimensions can be expressed as \cite{6},
\begin{equation}\label{M1}
\begin{split}
ds^2 = \chi(r) \left[-\left(\frac{r}{L}\right)^{2z} f(r) dt^2 + L^2 \frac{dr^2}{f(r)r^2} + r^2 d\Omega^2_{k, d-1} \right],
\end{split}
\end{equation}
where the function \(\chi(r)\) is defined as $ \chi(r) = \left(\frac{r}{r_F}\right)^{- \frac{2\theta}{d-1}}.$ For large values of \(r\), the function \(f(r)\) asymptotically approaches unity, i.e., \(f(r) \to 1\) as \(r \to \infty\). This metric has been proposed as a holographic dual to states in Lifshitz invariant field theories that exhibit hyperscaling violation. In the special case where \(z = 1\) and \(\theta = 0\), the geometry reduces to asymptotically Anti-de Sitter (AdS) spacetime. The parameter \(k\) determines the topology of spatial slices at constant \(t\) and \(r\), taking values: \( k = -1 \) for hyperbolic topology, \( k = 0 \) for planar topology, \( k = 1 \) for spherical topology. The analytic solutions describing these black holes are given by the metric above and generalize previously known solutions \cite{6},
\begin{equation}\label{eq1}
\begin{split}
f (r)=\frac{(d-2)^2 L^2}{r^2 (d-\theta +z-3)^2}-\frac{m}{r^{d-\theta +z-1}}+\frac{q^2}{r^{2 (d-\theta +z-2)}}+1
\end{split}
\end{equation}
In \cite{H2}, researchers extend the charged spherical black hole solutions with Lifshitz asymptotics (\(\theta \neq 0\)).
Also, in \cite {H3}, they study incorporating arbitrary Lifshitz scaling and hyperscaling violation parameters in charged black branes with non-trivial topology. with respect to \cite{6}, we have,
\begin{equation}\label{eq2}
\begin{split}
L^2=-\frac{r_0^{\frac{2 \theta }{d-1}} ((d-\theta +z-2) (d-\theta +z-1))}{2 \Lambda _0 r_F^{\frac{2 \theta }{d-1}}}
\end{split}
\end{equation}
and
\begin{equation}\label{eq3}
\begin{split}
\rho _3^2=\frac{r_F^{\frac{2 \theta }{d-1}} \left(2 q^2 (d-\theta -1) (d-\theta +z-3)\right)}{Z_0 L^{2 z} r_0^{\gamma  \lambda _3}}
\end{split}
\end{equation}
Considering $L$ as the bulk curvature radius (analogous to the AdS curvature radius) and $r_0$ as an arbitrary length scale, this solution holds under the conditions $d-\theta +z-3 >0$ and $\theta <d-1$. The corresponding black hole parameters include: ADM mass (M), Hawking temperature (T) are as follows,
\begin{equation}\label{eq4}
\begin{split}
M=\frac{m (d-\theta -1) L^{-z-1} r_F^{\theta } \omega _{k,d-1}}{16 \pi  G}
\end{split}
\end{equation}
\begin{equation}\label{eq5}
\begin{split}
T=\frac{r^z \left(\frac{(d-2)^2 L^2}{r^2 (d-\theta +z-3)}-\frac{q^2 (d-\theta +z-3)}{r^{2 (d-\theta +z-2)}}+d-\theta +z-1\right)}{4 \pi  L^{z+1}}
\end{split}
\end{equation}
The horizon radius $r_h$ is defined as the largest positive root of f(r) = 0. The mass M depends on the parameters $(m, L, r_F )$, where $m$ can be further expressed in terms of $r_h$, L, and q by solving $f(r_h, L, m, q) = 0$ for m. To establish a generalized Smarr relation and the first law, we express M as a function of the thermodynamic variables S and Q, along with the analog of the 'bulk pressure' $P:= - \Lambda_0 / 8 \pi G$ in the EMD theory \cite{4000}. Here, $\Lambda_0$ represents the 'bare' cosmological constant appearing in the Lagrangian. Due to its coupling to the dilaton, $\Lambda_0$ does not correspond to the pressure of a bulk perfect fluid in the conventional sense of the cosmological constant, yet we retain P as a useful theoretical parameter. Also, the entropy and charge of these models can be considered as,
\begin{equation}\label{eq6}
\begin{split}
S=\frac{r^{d-\theta -1} r_F^{\theta } \omega _{k,d-1}}{4 G}
\end{split}
\end{equation}
\begin{equation}\label{eq7}
\begin{split}
Q=\frac{\rho _3 Z_0 L^{z-1} \omega _{k,d-1} r_F^{\frac{\theta -2 \theta }{d-1}}}{16 \pi  G}
\end{split}
\end{equation}
The central charge \(C\) can be inferred from the proportionality constant in the grand canonical free energy expression: $F = M - T S - \Phi Q.$
Since mass, entropy, and charge scale with \(C\) (i.e., \(M, S, Q \propto C\)), its holographic dictionary entry can be determined by analyzing entropy expressions. The Bekenstein-Hawking entropy can be rewritten in terms of the scaling properties: $ S \propto C x^{d - \theta - 1},$
where \(x = r_h / L\), indicating as then the central charge \(C\)  defines as,
\begin{equation}\label{eq8}
\begin{split}
C=\frac{A L^{d-\theta -1} r_F^{\theta }}{G}
\end{split}
\end{equation}
Also, with respect to \cite{6}, we will have
\begin{equation}\label{eq9}
\begin{split}
\bar{Q}=Q L^{2-z} r_F^{\frac{\theta }{d-1}} r_0^{-\frac{\theta }{d-1}+z-1}
\end{split}
\end{equation}
The internal energy and temperature dictionary entries arise from symmetry considerations. The bulk quantities \(ML\) and \(TL\) are dimensionless, and should correspond to boundary quantities that remain invariant under,
\begin{equation}\label{eq10}
\begin{split}
\bar{E} R^{\frac{z-\frac{\theta}{d-1} }{1-\frac{\theta}{d-1}}}=L M
\end{split}
\end{equation}
and
\begin{equation}\label{eq11}
\begin{split}
\bar{T} R^{\frac{z-\frac{\theta}{d-1} }{1-\frac{\theta}{d-1}}}=L T
\end{split}
\end{equation}
 where $R$ denotes the radius of curvature. Its associated power depends on the selected black hole; it is also expressed as a special power, provided that scale scaling dimension invariance is preserved \cite{6}. So, we will have,
\begin{equation}\label{eq122}
\begin{split}
S=\overline{S}=\frac{C x^{d-\theta -1} \omega _{k,d-1}}{4 A}
\end{split}
\end{equation}
where $A$ is a constant. To derive the internal energy expression for the conformal field theory (CFT), we introduce dimensionless parameters that facilitate a more systematic formulation of thermodynamic quantities,
\begin{equation}\label{eq12}
\begin{split}
x=\frac{r_h}{L} 
\end{split}
\end{equation}
By applying the relationships established in Eqs. (\ref{eq4}), (\ref{eq6}), (\ref{eq8}), (\ref{eq10}), we obtain a comprehensive framework for evaluating thermodynamic properties of the CFT. So we will have,
\begin{equation}\label{eq13}
\begin{split}
&\bar{E}=-\frac{R^{\frac{z-\frac{\theta}{d-1} }{1-\frac{\theta}{d-1}}}\left(C (d-\theta -1) \omega _{k,d-1} x^{d-\theta +z-1}\right)}{16 \pi  A}\\
&\times \bigg(-\frac{128 \pi ^2 A^2 \bar{Q}^2 r_F^{2 \theta -\frac{2 \theta }{d-1}} x^{-2 (d-\theta +z-2)} r_0^{\gamma  \lambda _3+\frac{2 \theta }{d-1}-\frac{2 d z}{d-1}+\frac{2 z}{d-1}+\frac{2 d}{d-1}-\frac{2}{d-1}}}{C^2 Z_0 (d-\theta -1) (d-\theta +z-3) \omega _{k,d-1}^2}-\frac{(d-2)^2}{x^2 (d-\theta +z-3)^2}-1 \bigg)
\end{split}
\end{equation}
Also, temperature is calculated as, 
 \begin{equation}\label{eq14}
\begin{split}
&\bar{T}=\frac{x^z R^{\frac{z-\frac{\theta}{d-1} }{1-\frac{\theta}{d-1}}}}{4 \pi }\\
&\times \bigg(-\frac{128 \pi ^2 A^2 \bar{Q}^2 r_F^{2 \theta -\frac{2 \theta }{d-1}} x^{-2 (d-\theta +z-2)} r_0^{\gamma  \lambda _3+\frac{2 \theta }{d-1}-\frac{2 d z}{d-1}+\frac{2 z}{d-1}+\frac{2 d}{d-1}-\frac{2}{d-1}}}{C^2 Z_0 (d-\theta -1) \omega _{k,d-1}^2}+\frac{(d-2)^2}{x^2 (d-\theta +z-3)}+d-\theta +z-1\bigg)
\end{split}
\end{equation}
Furthermore, the critical points of the system can be determined using the following conditions,
$
\frac{\partial \tilde{T}}{\partial x} \Big|_{\overline{Q}, C} = 0,
$
$
\frac{\partial^2 \tilde{T}}{\partial x^2} \Big|_{\overline{Q}, C} = 0. $ Utilizing Eq. (\ref{eq14}), we extract the precise locations of these critical points \footnote{The solution obtained here for the case $(z=1, \theta=0)$ coincides with the solution of the charged AdS black hole (Eq. 4.2 in \cite{4'}). The main point, however, is that according to Eqs. (14) and (17), we obtain
\begin{equation*}\label{eq1}
\begin{split}
r_{h_{crt}}^2=\frac{(d-2)^2 (2-z)}{z (d-\theta+z-2)(d-\theta+z-1)}L^2.
\end{split}
\end{equation*}
Using the above equation and
$
C=\frac{AL^{d-\theta-1}r_F^{\theta}}{G},
$, we find
\begin{equation*}\label{eq2}
\begin{split}
r_{h_{crt}}^2=\frac{(d-2)^2 (2-z)}{z (d-\theta+z-2)(d-\theta+z-1)}\left(\frac{C G r_F^{-\theta}}{A}\right)^{\frac{2}{d-\theta-1}}.
\end{split}
\end{equation*}
Therefore, according to the above relation, the critical horizon radius depends on the central charge but remains independent of the electric charge.}.
\begin{equation}\label{eq15}
\begin{split}
x_{crt}^2=\frac{(d-2)^2 (2-z)}{z (d-\theta +z-2) (d-\theta +z-1)}
\end{split}
\end{equation}
As shown in Eq. (\ref{eq15}), $x_{crt}^2$ is independent of both the CFT framework and the central charge parameter. So, with respect to the above concepts, the $\bar{Q}^2{}_{crt}$ is calculated as follows,
\begin{equation}\label{eq17}
\begin{split}
&\bar{Q}^2{}_{crt}= C^2 z^3 Z_0 \left(-\frac{(d-2)^2 (z-2)}{z (d-\theta +z-2) (d-\theta +z-1)}\right)^{d-\theta +z} \omega _{k,d-1}^2 r_F^{-\frac{2 (d-2) \theta }{d-1}} r_0^{-\gamma  \lambda _3-\frac{2 \theta }{d-1}+2 z-2}\\
&\times \frac{(d-\theta -1) (d-\theta +z-2)^2 (d-\theta +z-1)^3}{128 \pi ^2 A^2 (d-2)^4 (z-2)^2 (d-\theta +z-3) (2 d-2 (\theta +2)+z)}
\end{split}
\end{equation}
Finally, employing the formulations in equations (\ref{eq6}), (\ref{eq13}), and (\ref{eq14}), we determine the Helmholtz free energy, which serves as a key thermodynamic potential for characterizing the stability and equilibrium properties of the CFT.
\begin{equation}\label{eq18}
\begin{split}
\bar{F}=\bar{E}-S \bar{T}
\end{split}
\end{equation}
Also, we can obtain,
 \begin{equation}\label{eq19}
\begin{split}
\bar{F}=\frac{C R^{\frac{z-\frac{\theta}{d-1} }{1-\frac{\theta}{d-1}}} \omega _{k,d-1} x^{d-\theta +z-1} \left(\frac{128 \pi ^2 A^2 \bar{Q}^2 (2 d-2 \theta +z-4) r_F^{\frac{2 (d-2) \theta }{d-1}} x^{-2 (d-\theta +z-2)} r_0^{\gamma  \lambda _3+\frac{2 \theta }{d-1}-2 z+2}}{c^2 Z_0 (d-\theta -1) (d-\theta +z-3) \omega _{k,d-1}^2}-\frac{(d-2)^2 (z-2)}{x^2 (d-\theta +z-3)^2}-z\right)}{16 \pi  A}
\end{split}
\end{equation}
Since our objective is to investigate the thermodynamic properties, phase transitions, and stability of the black hole within the holographic framework, it is essential to first determine the energy of the black hole in the dual conformal field theory (CFT) space. The energy in the CFT provides crucial insights into how gravitational phenomena in the bulk correspond to thermal properties in the boundary field theory, thereby enabling a deeper understanding of holographic duality. Furthermore, to analyze the critical behavior of the black hole within the CFT space, particularly the existence and characterization of critical points, it is necessary to derive the temperature of the black hole as measured in this dual description. The temperature plays a fundamental role in defining the thermodynamic state variables and allows us to probe the phase structure of the system. From equation~(17), one can deduce that when the dynamical critical exponent \( z \) satisfies
$
z > 2,
$
and the combination of parameters satisfies
$
d - \theta + z - 2 > 0,
$
No critical points appear. This implies that under these conditions, the black hole does not undergo phase transitions, reflecting a stable thermodynamic phase in the holographic theory. In addition to the above, to systematically investigate both first-order and second-order phase transitions, we consider the Helmholtz free energy \( \bar{F} \) of the black hole. By calculating \( \bar{F} \), we can identify the thermodynamic stability of different phases and characterize the nature of the phase transitions. In particular, discontinuities or non-analytic behaviors in the free energy and its derivatives serve as indicators of first-order or continuous phase transitions, respectively. This approach provides a comprehensive framework to understand the thermodynamics of black holes in holographic settings and their corresponding dual field theories.\\\\
According to Figs. 1-4, we can examine the first and second order phase transitions for a black hole in the CFT thermodynamic state. In Figs 1 and 2, we analyzed the scenario with a fixed central charge, examining the electric charge both above and below its critical values. For $\overline{Q}<\overline{ Q}_{crt}$, the free energy exhibits a swallowtail behavior, indicating a first-order phase transition between two thermodynamically stable branches. The horizontal branch with lower entropy corresponds to the small black hole, and the vertical branch with higher entropy corresponds to the large black hole. In contrast, the middle branch connecting these two branches is unstable due to its negative heat capacity. Also, for $\overline{Q}=\overline{Q}_{crt}$, we have a second-order phase transition, and the small stable black hole directly transforms into a large stable black hole. However, when $\overline{Q}>\overline{Q}_{crt}$, no phase transition takes place. In Figs 3 and 4, we examine the charge $\overline{Q}$ fixed and different values of central charge above and below the critical points. For $C>C_{crt}$, there is a first-order phase transition, for $C=C_{crt}$, there is a second-order phase transition, and for $C<C_{crt}$, there is no phase transition.

  \begin{figure}[H]
 \begin{center}
 \subfigure[]{
 \includegraphics[height=5cm,width=7cm]{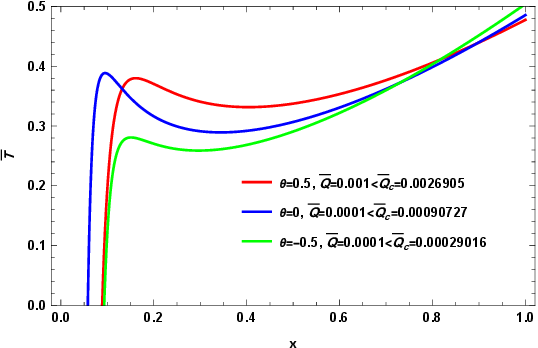}
 \label{F2a}}
 \subfigure[]{
 \includegraphics[height=5cm,width=7cm]{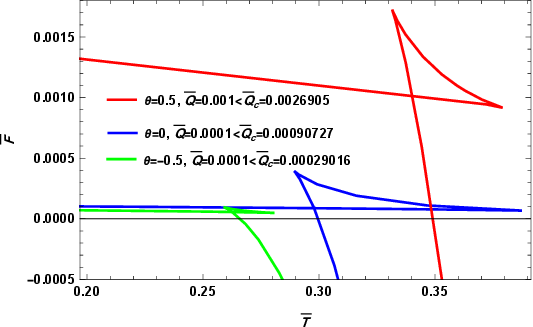}
 \label{F1b}}
  \caption{\small{Each plot illustrates the behavior of the system for three distinct values of the parameter \(\theta\), while all other parameters are held constant to isolate the effects of varying \(\theta\). Specifically, the fixed parameters are set as follows: \(k = 1\)
; \(Z_0 = 1\)
; \(\gamma = 1\)
; \(\lambda_3 = 1\)
; \(R = 1\)
; \(r_0 = 1\)
; \(r_F = 1\)
; \(A = 1\)
; \(\omega_{k,d-1} = 1\)
; \(C = 2\)
; \(z = \frac{3}{2}\)
; \(d = 4\)
The plots are organized into two panels: Panel (a): This plot shows the variation of temperature as a function of the spatial coordinate \(x\). By comparing the curves corresponding to different values of \(\theta\), one can observe how the temperature profile evolves with respect to changes in \(\theta\), under the influence of the fixed background parameters. Panel (b): This plot presents the free energy as a function of temperature. It provides insight into the thermodynamic stability and phase structure of the system. The distinct curves for each value of \(\theta\) reveal how the free energy landscape is modified by \(\theta\), potentially indicating transitions or critical behavior.}}
 \label{ّF2F1}
\end{center}
 \end{figure}
 
  \begin{figure}[H]
 \begin{center}
 \subfigure[]{
 \includegraphics[height=5cm,width=7cm]{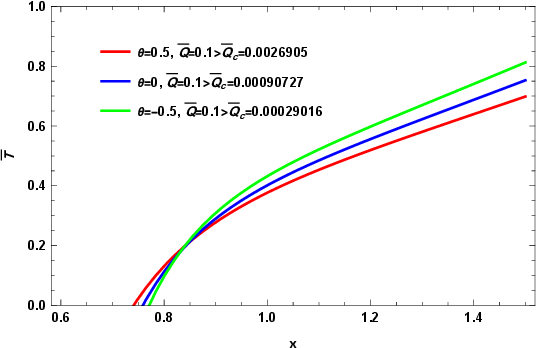}
 \label{F6a}}
 \subfigure[]{
 \includegraphics[height=5cm,width=7cm]{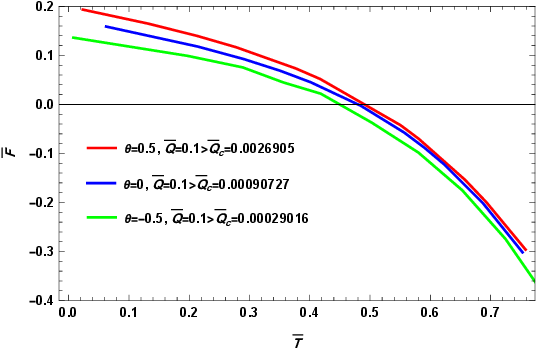}
 \label{F5b}}
  \caption{\small{Each plot illustrates the behavior of the system for three distinct values of the parameter \(\theta\), allowing for a comparative analysis of its influence on the physical quantities under consideration. All other parameters are held fixed to ensure that variations in the plots arise solely due to changes in \(\theta\). The fixed parameters are specified as follows:
\(k = 1\)
; \(Z_0 = 1\)
; \(\gamma = 1\)
; \(\lambda_3 = 1\)
; \(R = 1\)
; \(r_0 = 1\)
; \(r_F = 1\)
; \(A = 1\)
; \(\omega_{k,d-1} = 1\)
; \(C = 2\)
; \(z = \frac{3}{2}\)
; \(d = 4\)
The results are presented in two separate panels:
Panel (a): Temperature vs. \(x\):  This plot displays the temperature profile as a function of the spatial coordinate \(x\). By examining the curves corresponding to different values of \(\theta\), one can observe how the temperature distribution evolves across space. The comparison highlights the role of \(\theta\) in shaping the thermal behavior of the system, while the influence of other parameters remains constant. Panel (b): Free Energy vs. Temperature:  This plot illustrates the relationship between free energy and temperature for the same set of \(\theta\) values. It provides insight into the thermodynamic properties and stability of the system. Differences in the free energy curves reflect how variations in \(\theta\) affect the system’s energetics, potentially revealing phase transitions or critical points.}}
 \label{F6F5}
\end{center}
 \end{figure}
 
   \begin{figure}[H]
 \begin{center}
 \subfigure[]{
 \includegraphics[height=5cm,width=7cm]{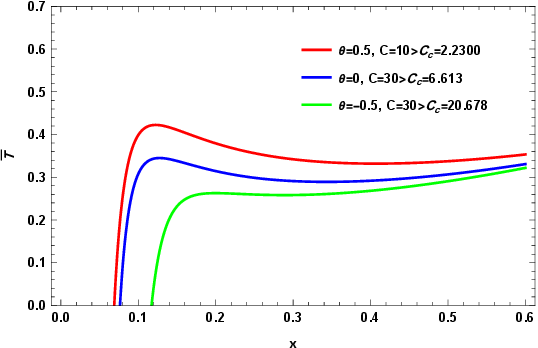}
 \label{F4a}}
 \subfigure[]{
 \includegraphics[height=5cm,width=7cm]{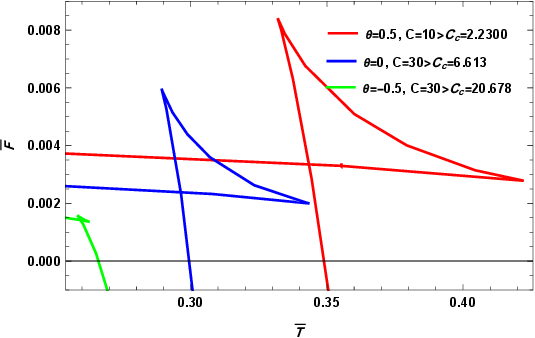}
 \label{F3b}}
  \caption{\small{Each plot illustrates the behavior of the system for three distinct values of the parameter \(\theta\), enabling a comparative analysis of its impact on the physical observables. To isolate the effects of \(\theta\), all other parameters are held constant throughout the analysis. The fixed parameters are chosen as follows:
\(k = 1\)
; \(Z_0 = 1\)
; \(\gamma = 1\)
; \(\lambda_3 = 1\)
; \(R = 1\)
; \(r_0 = 1\)
; \(r_F = 1\)
; \(A = 1\)
; \(\omega_{k,d-1} = 1\)
; \(\overline{Q} = 0.003\)
; \(z = \frac{3}{2}\)
; \(d = 4\)
The results are presented in two panels, each highlighting a different physical relationship: Panel (a): Temperature vs. \(x\):  This plot depicts the temperature profile as a function of the spatial coordinate \(x\). The curves corresponding to three distinct values of \(\theta\) reveal how the temperature distribution varies across space. By keeping all other parameters fixed, the influence of \(\theta\) on the thermal structure of the system can be clearly discerned. This comparison provides insight into how anisotropic or scaling effects governed by \(\theta\) manifest in the temperature field. Panel (b): Free Energy vs. Temperature: This plot shows the dependence of the free energy on temperature for the same set of \(\theta\) values. It serves to characterize the thermodynamic behavior of the system, including potential phase transitions or stability regimes. Variations in the free energy curves reflect how changes in \(\theta\) affect the system’s energetic landscape, offering clues about the underlying microscopic or geometric structure.}}
 \label{F4F3}
\end{center}
 \end{figure}

  \begin{figure}[H]
 \begin{center}
 \subfigure[]{
 \includegraphics[height=5cm,width=7cm]{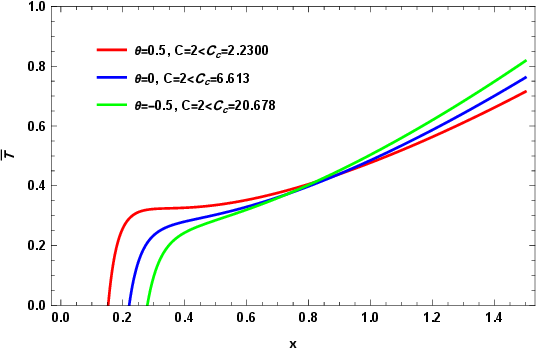}
 \label{F8a}}
 \subfigure[]{
 \includegraphics[height=5cm,width=7cm]{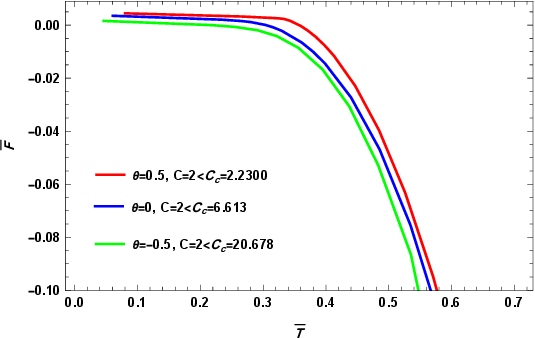}
 \label{F7b}}
  \caption{\small{Each plot illustrates the behavior of the system for three distinct values of the parameter \(\theta\), allowing for a focused investigation into its influence on the physical observables. To ensure that the observed variations are solely attributable to changes in \(\theta\), all other parameters are held fixed throughout the analysis. The chosen fixed values are:
\(k = 1\)
; \(Z_0 = 1\)
; \(\gamma = 1\)
; \(\lambda_3 = 1\)
; \(R = 1\)
; \(r_0 = 1\)
; \(r_F = 1\)
; \(A = 1\)
; \(\omega_{k,d-1} = 1\)
; \(\overline{Q} = 0.003\)
; \(z = \frac{3}{2}\)
; \(d = 4\)
The results are presented in two panels, each capturing a distinct thermodynamic or spatial relationship: Panel (a): Temperature vs. Spatial Coordinate \(x\): This plot displays the temperature profile as a function of the spatial coordinate \(x\) for three different values of \(\theta\). By examining these curves, one can observe how the temperature distribution evolves across space in response to changes in \(\theta\). The fixed background parameters ensure that the influence of \(\theta\) is isolated, highlighting its role in modifying the thermal structure of the system. Such variations may reflect underlying changes in the geometry, scaling behavior, or anisotropic features governed by \(\theta\). Panel (b): Free Energy vs. Temperature:  
  This plot presents the free energy as a function of temperature, again for three distinct values of \(\theta\). The comparison of these curves provides insight into the thermodynamic stability and phase behavior of the system. Differences in the free energy profiles suggest that \(\theta\) plays a significant role in shaping the energetic landscape, potentially indicating the presence of critical points, phase transitions, or shifts in equilibrium configurations.}}
 \label{F8F7}
\end{center}
 \end{figure}
\section{Stability}
To gain deeper insights into the thermodynamic properties and stability of the dirty black hole, we examine the specific heat at constant parameters. The specific heat can be computed via the thermodynamic relation,
\begin{equation}\label{C1}
\widetilde{C }= \overline{T} \left( \frac{\partial S}{\partial \overline{T}} \right) = \overline{T} \left( \frac{\partial S}{\partial x} \right) \left( \frac{\partial x}{\partial \overline{T}} \right).
\end{equation}
So, we can calculate the $\widetilde{C }$ using the Hawking temperature from Eq. (\ref{eq14}) and entropy from Eq. (\ref{eq6}),
\begin{equation}\label{C2}
\widetilde{C }=\frac{C (d-\theta -1) x^{d-\theta +1} \omega _{k,d-1} \left(-\frac{128 \pi ^2 A^2 \overline{Q}^2 r_F^{\frac{2 (d-2) \theta }{d-1}} x^{-2 (d-\theta +z-2)} r_0^{\gamma  \lambda _3+\frac{2 \theta }{d-1}-2 z+2}}{C^2 Z_0 (d-\theta -1) \omega _{k,d-1}^2}+\frac{(d-2)^2}{x^2 (d-\theta +z-3)}+d-\theta +z-1\right)}{4 A \left(\frac{128 \pi ^2 A^2 \overline{Q}^2 (2 d-2 \theta +z-4) r_F^{\frac{2 (d-2) \theta }{d-1}} x^{-2 (d-\theta +z-3)} r_0^{\gamma  \lambda _3+\frac{2 \theta }{d-1}-2 z+2}}{C^2 Z_0 (d-\theta -1) \omega _{k,d-1}^2}+x^2 z (d-\theta +z-1)+\frac{(d-2)^2 (z-2)}{d-\theta +z-3}\right)}
\end{equation}
This formulation of the specific heat reveals several key features of the black hole's thermodynamic behavior:
A second-order phase transition occurs at the points where the specific heat diverges. This signifies a fundamental change in the thermal properties of the black hole. When in some region, the specific heat is positive, indicating thermal stability, and in some other region, the specific heat becomes negative, signifying thermal instability.  
These findings are consistent with our topological analysis, where configurations with positive winding numbers correspond to thermally stable states, while negative winding numbers indicate instability.
The specific heat demonstrates that black holes experience enhanced thermal stability. This dependence implies that the stability of the system strengthens significantly as the black hole size increases.
\begin{figure}[]
 \begin{center}
 \subfigure[]{
 \includegraphics[height=6cm,width=7.5cm]{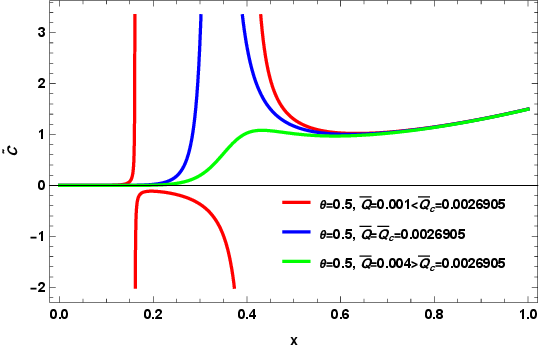}
 \label{100a}}
 \subfigure[]{
 \includegraphics[height=6cm,width=7.5cm]{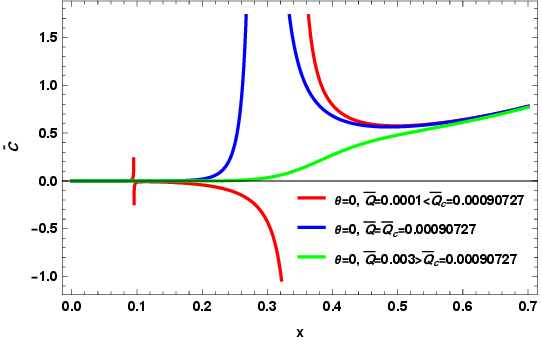}
 \label{100b}}
 \subfigure[]{
 \includegraphics[height=6cm,width=7.5cm]{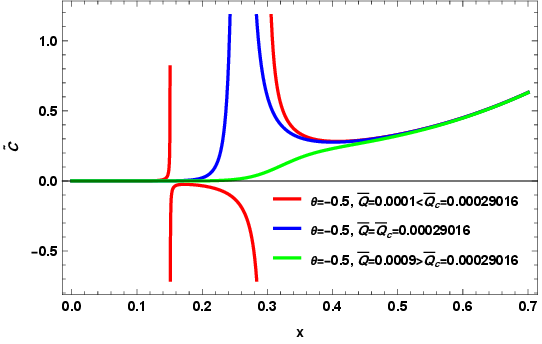}
 \label{100c}}
  \caption{\small{The stability plot, which presents the heat capacity \(\widetilde{C}\) as a function of the spatial coordinate \(x\), provides valuable insight into the thermodynamic behavior of the system. This plot is constructed under a specific set of fixed parameters to isolate the effects of spatial variation and ensure consistency across different scenarios. The parameters are chosen as follows:
\(k = 1\)
; \(Z_0 = 1\)
; \(\gamma = 1\)
; \(\lambda_3 = 1\)
; \(R = 1\)
; \(r_0 = 1\)
; \(r_F = 1\)
; \(A = 1\)
; \(\omega_{k,d-1} = 1\)
; \(C = 2\)
; \(z = \frac{3}{2}\)
; \(d = 4\)
In this context, the heat capacity \(\widetilde{C}\) serves as a diagnostic tool for assessing the local thermodynamic stability of the system. Positive values of \(\widetilde{C}\) typically indicate stable configurations, whereas negative values may signal instability or the presence of phase transitions. By plotting \(\widetilde{C}\) against \(x\), one can identify regions where the system exhibits stable or unstable behavior, and observe how these regions are influenced by the underlying geometry and scaling properties encoded in the fixed parameters. This analysis is particularly relevant in systems with spatially dependent thermodynamic quantities, where local variations in stability can have significant implications for the global behavior of the system. The choice of fixed parameters ensures that the observed features in the \(\widetilde{C}-x\) plot are intrinsic to the spatial structure and not artifacts of varying background conditions.}}
 \label{ّfigP1}
\end{center}
 \end{figure}
 \begin{figure}[]
 \begin{center}
 \subfigure[]{
 \includegraphics[height=6cm,width=7.5cm]{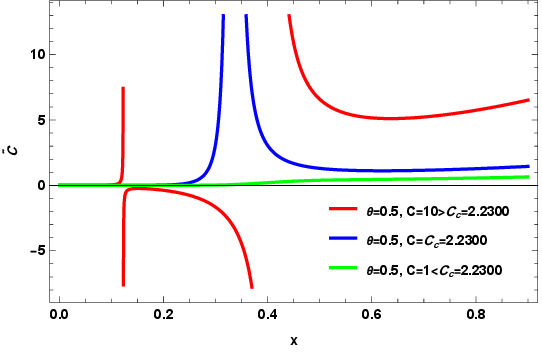}
 \label{200a}}
 \subfigure[]{
 \includegraphics[height=6cm,width=7.5cm]{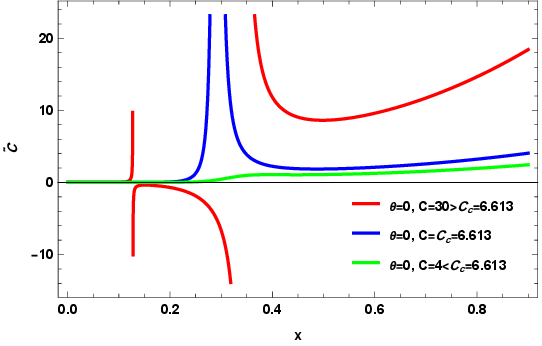}
 \label{200b}}
 \subfigure[]{
 \includegraphics[height=6cm,width=7.5cm]{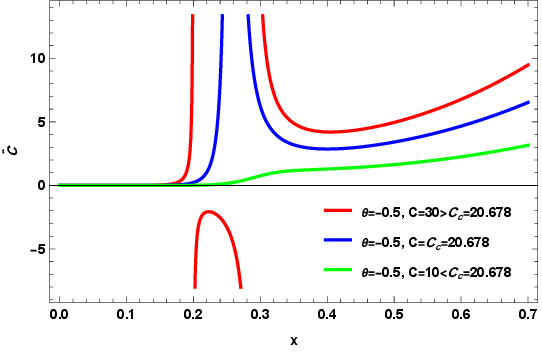}
 \label{200c}}
  \caption{\small{The stability plot, denoted as \(\widetilde{C}\) versus \(x\), illustrates the behavior of the heat capacity across spatial coordinates and serves as a key indicator of the system’s thermodynamic stability. In this analysis, the plot is generated under a controlled set of fixed parameters to ensure that the observed variations in heat capacity are solely due to spatial effects and not influenced by changes in other model inputs. The parameters are fixed as follows:
\(k = 1\)
; \(Z_0 = 1\)
; \(\gamma = 1\)
; \(\lambda_3 = 1\)
; \(R = 1\)
; \(r_0 = 1\)
; \(r_F = 1\)
; \(A = 1\)
; \(\omega_{k,d-1} = 1\)
; \(\overline{Q} = 0.003\)
; \(z = \frac{3}{2}\)
; \(d = 4\)
In this context, the heat capacity \(\widetilde{C}\) is a crucial thermodynamic quantity that reflects the system’s response to temperature fluctuations. Positive values of \(\widetilde{C}\) typically correspond to stable thermodynamic phases, where the system absorbs heat without undergoing abrupt changes. Conversely, negative values may signal instability, indicating regions where small perturbations can lead to significant shifts in the system’s configuration or phase. By plotting \(\widetilde{C}\) as a function of the spatial coordinate \(x\), the diagram reveals how stability varies throughout the system. This spatial dependence may arise from geometric, scaling, or field-theoretic features embedded in the model, particularly influenced by the dynamical exponent \(z\) and the dimensionality \(d\). The inclusion of a small but finite charge parameter \(\overline{Q} = 0.003\) further enriches the thermodynamic structure, potentially introducing subtle effects related to charge density or electromagnetic interactions.}}
 \label{ّfigP2}
\end{center}
 \end{figure}
As shown in Fig. 5, for different values of the free parameter—particularly $\overline{Q}$—the system exhibits instability or negative values only within a very small range where $\overline{Q} < \overline{Q}_c$. However, at all other points, it maintains a positive value, indicating stability. This behavior is clearly visible in Fig. 5. 
Conversely, the parameter $C$ exhibits the opposite trend. Specifically, except for a narrow range where $C > C_{crt}$ Fig. 6, the system demonstrates instability with negative values. Yet, at most points, it remains stable. 
This contrast in the structural behavior of $\overline{Q}$ and $C$ is well illustrated in the study of the thermodynamic topology of black holes, particularly in the context of first-order and second-order phase transitions.
\section{Thermodynamic Topology}
The integration of topology into black hole thermodynamics provides a powerful framework for analyzing phase transitions. By assigning specific topological indices to critical points in the phase space, this method uncovers novel classifications and behaviors that conventional thermodynamic analyses might overlook. The foundation of this approach lies in the generalized Helmholtz free energy, which serves as a key tool for evaluating stability and transition dynamics. The generalized Helmholtz free energy, an extension of the traditional Helmholtz free energy, incorporates off-equilibrium effects and is defined as,
\begin{equation}\label{T1}
F(S, V, Y) = U(S, V, Y) - T S.
\end{equation}
Here, \(S\) represents entropy, \(V\) denotes volume, \(Y\) accounts for additional thermodynamic degrees of freedom, \(T\) is the temperature, and \(U\) is the internal energy. Within the framework of conformal field theory (CFT) thermodynamics, the phase behavior of black holes can be characterized using the Helmholtz free energy expression $\overline{ F}= \overline{E} - \overline{T}\overline{S}.$ This formulation encapsulates the relationship between energy, temperature, and entropy in the thermodynamic analysis of black hole systems. A deeper understanding of thermodynamic topology requires analyzing the modified Helmholtz free energy function, which is given by \cite{a19,a20,20a,21a,22a,23a,24a,25a,26a,27a,28a,29a,31a,33a,34a,35a,37a,38',38a,38b,38c,39a,40a,41a,42a,43a,44a,44c,44d,44e,44f,44g,44h,44i,44j,44k,44l,44m},
\begin{equation}\label{T2}
\mathcal{F} = M - \frac{S}{\tau}.
\end{equation}
Here, \(M\) represents the black hole mass, while \(\tau\) corresponds to the inverse temperature period in the Euclidean formulation. To investigate the phase structure of black holes using conformal field theory (CFT) thermodynamics, we reformulate the generalized Helmholtz energy within the CFT framework as \cite{44h},
\begin{equation}\label{T3}
\mathcal{\overline{ F}}= \overline{ E} - \frac{\overline{S}}{\overline{\tau}}.
\end{equation}
In this formulation, the Helmholtz energy attains an on-shell condition exclusively when the time parameter satisfies $\overline{\tau} =\overline{ \tau}_H = \frac{1}{\overline{T}_h}.$ This constraint ensures consistency with holographic thermodynamics, reinforcing the connection between black hole phase transitions and the underlying conformal field dynamics. Examining the topological features of this function reveals a vector structure,
\begin{equation}\label{T4}
\boldsymbol{\phi} = (\phi^{r_h}, \phi^\Theta),
\end{equation}
where \( \phi^{r_h}\) is derived from the derivative of \(F\) with respect to \(r\), and \( \phi^\Theta \) contains angular dependencies influencing phase stability. This vector can be reformulated as,
\begin{equation}\label{T5}
\boldsymbol{\phi} = ||\boldsymbol{\phi}|| e^{i\Theta}.
\end{equation}
Applying topological field theory, the normalized vector \(n^a\) is constructed as,
\begin{equation}\label{T6}
n^a = \frac{\phi^a}{|\boldsymbol{\phi}|},
\end{equation}
This allows defining the antisymmetric superpotential governing the thermodynamic phase transitions,
\begin{equation}\label{T7}
V^{\mu\nu} = \frac{1}{2\pi} \epsilon^{\mu\nu\rho} \epsilon^{ab} n^a \partial_\rho n^b.
\end{equation}
According to Duan's method, for the general superpotential given above, the following current can be defined
\begin{equation}\label{T8}
j^\mu = \frac{1}{2\pi} \epsilon^{\mu\nu\rho} \epsilon^{ab} \partial_\nu n^a \partial_\rho n^b.
\end{equation}
This topological current remains conserved due to Noether’s theorem,
\begin{equation}\label{T9}
\partial_\mu j^\mu = 0.
\end{equation}
 In that case, the component $ j^0$ will be equivalent to the Noether charge density, which can be integrated to calculate the topological charge:
\begin{equation*}
Q=\int_{\Omega}j^0 d^{2}x.
\end{equation*}
Now, if we replace the potential $ \phi$ with $ n^a$ in the original equation (instead of the superpotential), the result will be:
\begin{equation*}
j^{\mu}=\frac{1}{2\pi}\epsilon^{\mu\nu\rho}\epsilon_{ab}\frac{\partial}{\partial\phi^{c}}
\left(\frac{\phi^a}{||\phi||^2}\right)\partial_{\nu}\phi^{c}\partial_{\rho}\phi^{b}.
\end{equation*}
Using the derivative properties, together with the concept of the Jacobian tensor and the two-dimensional Laplacian Green function in the mapping space $ \phi$, which are given below, respectively:
$$
\left\{
\begin{aligned}
& \frac{\partial \ln||\phi||}{\partial\phi^a}=\frac{\phi^a}{||\phi||^2}, \\
& \epsilon^{ab}J^{\mu}\left(\frac{\phi}{x}\right)=\epsilon^{\mu\nu\rho} \, 
\partial_{\nu}\phi^a \, \partial_{\rho}\phi^b, \\
& \frac{\partial}{\partial\phi^a}\frac{\partial}{\partial\phi^a} \ln||\phi||=2\pi\delta(\phi),
\end{aligned}
\right.
$$
the topological current take the following simplified form:
\begin{equation*}
j^{\mu}=\delta^{2}(\phi)J^{\mu}\left(\frac{\phi}{x}\right). \label{juu}
\end{equation*}
Now, considering the properties of the delta function $ \delta$, it is well known that $ j^\mu$ is nonzero only at the zero points of the function $ \phi^a$. In particular, for the zeroth component we obtain:
\begin{equation}
j^{0}=\delta^{2}(\phi)J^{0}\left(\frac{\phi}{x}\right).
\end{equation}
Substituting this into the expression for the topological charge, we have:
\begin{equation}
Q=\int_{\Sigma}\delta^{2}(\phi)J^{0}\left(\frac{\phi}{x}\right)d^2x. \label{qcharge}
\end{equation}
As stated above, since the delta function ensures nonzero contributions only at the zeros of $ \phi^a$, it follows that the topological charge $Q$ is also nonzero only at these points. Referring back to the definition of $ \phi^a$, we clearly see that the zeros of $ \phi^a$ correspond to the zeros of $ \partial_r H$, i.e., the zeros of the derivative of the effective potential, which determine the locations of the photon spheres. Therefore, we can attribute a topological charge to each photon sphere.
By analyzing zero points of $(j^\mu)$, one can determine the winding number,
\begin{equation}\label{T10}
W = \sum_{i=1}^{n} \beta_i \eta_i = \sum_{i=1}^{n} \omega_i.
\end{equation}
This index captures the intricate topological features of black hole phase transitions by quantifying the mapping of thermodynamic variables in phase space. The Hopf index \( \beta_i \) measures vector linking behaviors, while \( \omega_i \) encapsulates rotational properties of the field structure. Through these topological invariants, black hole thermodynamics acquires a more refined classification scheme, offering deeper insights into its fundamental nature.\\Given these premises and considering Eqs. (\ref{eq13}) and (\ref{T3}), the general form of the quantities required for the topological study of the model will be as follows,
\begin{equation}\label{F1}
\begin{split}
&\mathcal{\overline{F}} =-\frac{R^{\frac{z-\frac{\theta}{d-1} }{1-\frac{\theta}{d-1}}} \omega_{k ,d -1} (-1-\frac{(d -2)^{2}}{x^{2} (d -\theta +z -3)^{2}}-\frac{128 r_{F}^{2 \theta -\frac{2 \theta}{d -1}} r_{0}^{\frac{-2 z d +2 d +2 \theta +2 z -2}{d -1}+\lambda_{3} \gamma} \overline{Q}^{2} x^{4-2 d -2 z +2 \theta} \pi^{2} A^{2}}{\omega_{k ,d -1}^{2} (d -\theta +z -3) (d -\theta -1) C^{2} Z_{0}}) (d -\theta -1) x^{d -\theta +z -1} C}{16 A \pi}\\&-\frac{x^{d -\theta -1} C \omega_{k ,d -1}}{4 A \tau}
\end{split}
\end{equation}
We can calculate the $\phi^{x}$ and $\phi^{\Theta}$ with respect to (\ref{T4}) as follows,
\begin{equation}\label{F2}
\begin{split}
&\phi^{x}=\bigg(8 \bigg[x^{3} A^{2} \overline{Q}^{2} (-d +\theta -z +3) r_{F}^{\frac{2 \theta  (d -2)}{d -1}} \pi^{2} (-2 x^{z -1} x^{3-2 d -2 z +2 \theta} (2-d -z +\theta ) R^{\frac{z-\frac{\theta}{d-1} }{1-\frac{\theta}{d-1}}}\\&+x^{z -2} x^{4-2 d -2 z +2 \theta} R^{\frac{z-\frac{\theta}{d-1} }{1-\frac{\theta}{d-1}}} (-d +\theta -z +1)) \tau  r_{0}^{\frac{\gamma  (d -1) \lambda_{3}+(-2 z +2) d +2 z +2 \theta -2}{d -1}}\\&+\bigg[(-d +\theta +1) \omega_{k ,d -1}^{2} Z_{0} (2 x^{z -1} \tau  (d -2)^{2} R^{\frac{z-\frac{\theta}{d-1} }{1-\frac{\theta}{d-1}}}+x (R^{\frac{z-\frac{\theta}{d-1} }{1-\frac{\theta}{d-1}}} \tau  (-d +\theta -z +1) ((-d +\theta -z +3)^{2} x^{2}+(d -2)^{2}) x^{z -2}\\&+4 \pi  (-d +\theta -z +3)^{2})) C^{2}\bigg]\bigg/\bigg(128\bigg)\bigg] x^{d -\theta}\bigg)\bigg/\bigg(C \omega_{k ,d -1} A \pi  (-d +\theta -z +3)^{2} Z_{0} x^{3} \tau\bigg)
\end{split}
\end{equation}
\begin{equation}\label{F3}
\phi^{\Theta}=- \cot \Theta \csc \Theta
\end{equation}
Also, we can calculate the $\tau$ as follows,
\begin{equation}\label{F4}
\begin{split}
&\overline{\tau} =-\bigg[4 R^{\frac{z-\frac{\theta}{d-1} }{1-\frac{\theta}{d-1}}} (d -\theta -1) (d -\theta +z -3) \omega_{k ,d -1}^{2} Z_{0} x^{2} \pi  C^{2}\bigg]\times\bigg[128 A^{2} x^{6-2 d -z +2 \theta} \overline{Q}^{2} r_{F}^{\frac{2 \theta  (d -2)}{d -1}} \pi^{2} (d -\theta +z -3)\\&\times r_{0}^{\frac{\gamma  (d -1) \lambda_{3}+(-2 z +2) d +2 z +2 \theta -2}{d -1}}-((d -\theta +z -1) (d -\theta +z -3) x^{2+z}+x^{z} (d -2)^{2}) C^{2} (d -\theta -1) Z_{0} \omega_{k ,d -1}^{2}\bigg]^{-1}
\end{split}
\end{equation}
In the classical or traditional framework, the standard approach to studying phase behavior involves identifying the critical point (critical boundary), which, depending on the structure of the equations derived from the model, typically either critical temperature or critical pressure is used. Once this boundary is established, phase behavior is examined on both sides by analyzing one of the free energy forms (Helmholtz or Gibbs) to determine whether the relevant function remains continuous or exhibits discontinuities.\\ 
Generally, due to the intrinsic relationship between temperature, pressure, and free energy, below the critical temperature, we observe sharp points and discontinuities, often forming a swallowtail structure, which indicates a first-order phase transition. Conversely, above this boundary, the system exhibits a smooth and continuous curve, signifying a second-order phase transition, where changes occur seamlessly without forming an unstable minimum energy state.
Correspondingly, in the topological analysis of black hole phase behavior, different black hole configurations (small, intermediate, and large) manifest as zero points characterized by their topological charge. 
In a first-order phase transition, the unstable intermediate black hole typically exhibits a topological charge of -1, while the stable small and large black holes display charges of +1. Conversely, in a second-order phase transition, only one black hole state emerges, generally carrying a topological charge of +1. This distinct behavioral difference is also clearly reflected in the $\tau$ diagram.\\
In this study, instead of considering pressure or temperature as the critical parameter, we analyze the critical boundary in terms of central charge (C) or charge (Q) within the framework of CFT. By doing so, we aim to explore how phase behavior varies with these parameters, essentially playing a more influential role in the CFT structure.\\
Considering the critical values obtained in Figs. 1-4 and considering the chosen values for $\Theta$, we will carry out the topological study in three parts.
\subsection{$\theta=0.5$, $C_{crt}=2.23$ }
Here, Given the central charge, we will witness a first-order phase transition at $C>C_{crt}$, and we will witness a second-order phase transition at $C<C_{crt}$.\\\\
\begin{figure}[H]
 \begin{center}
 \subfigure[]{
 \includegraphics[height=6cm,width=7cm]{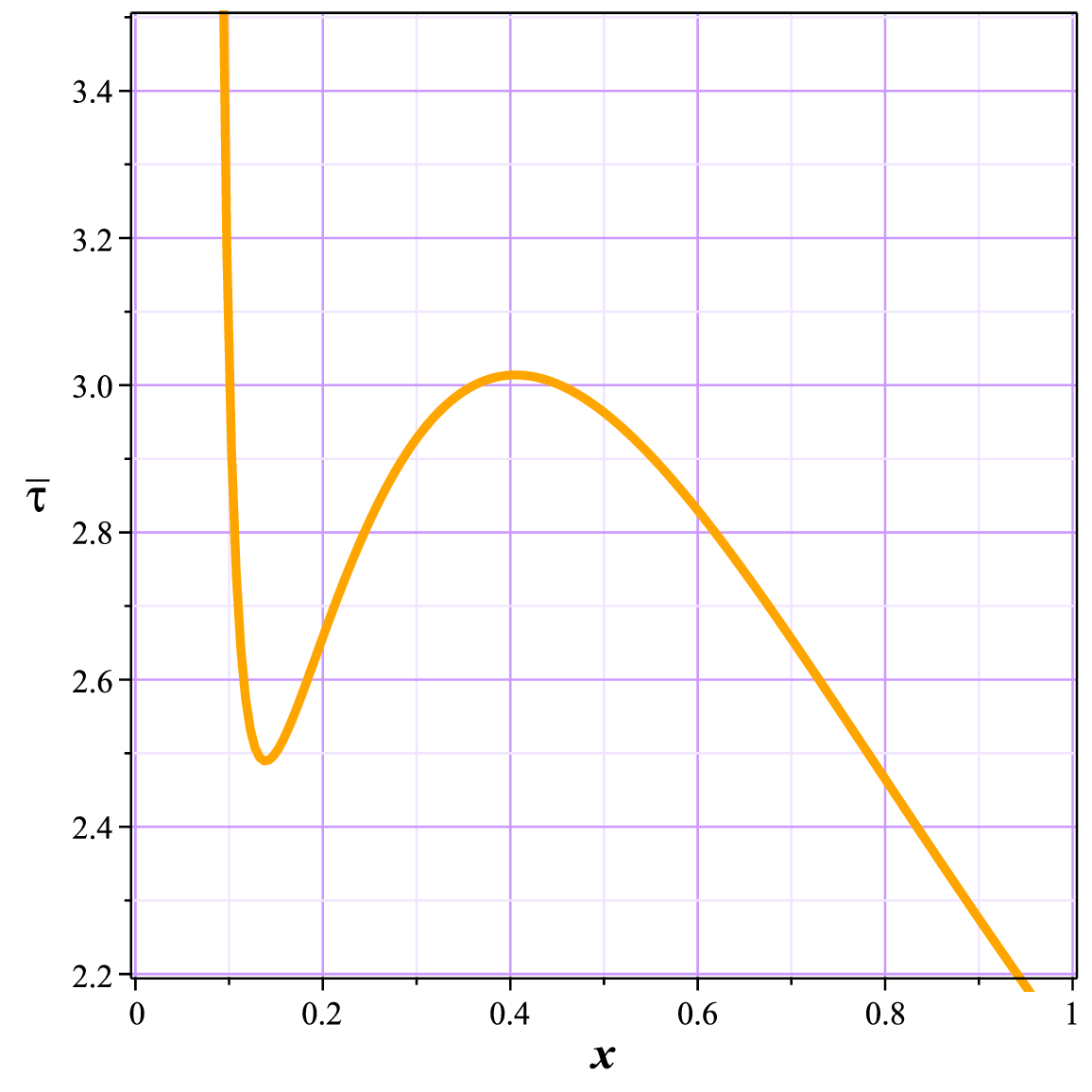}
 \label{300a}}
 \subfigure[]{
 \includegraphics[height=6cm,width=7cm]{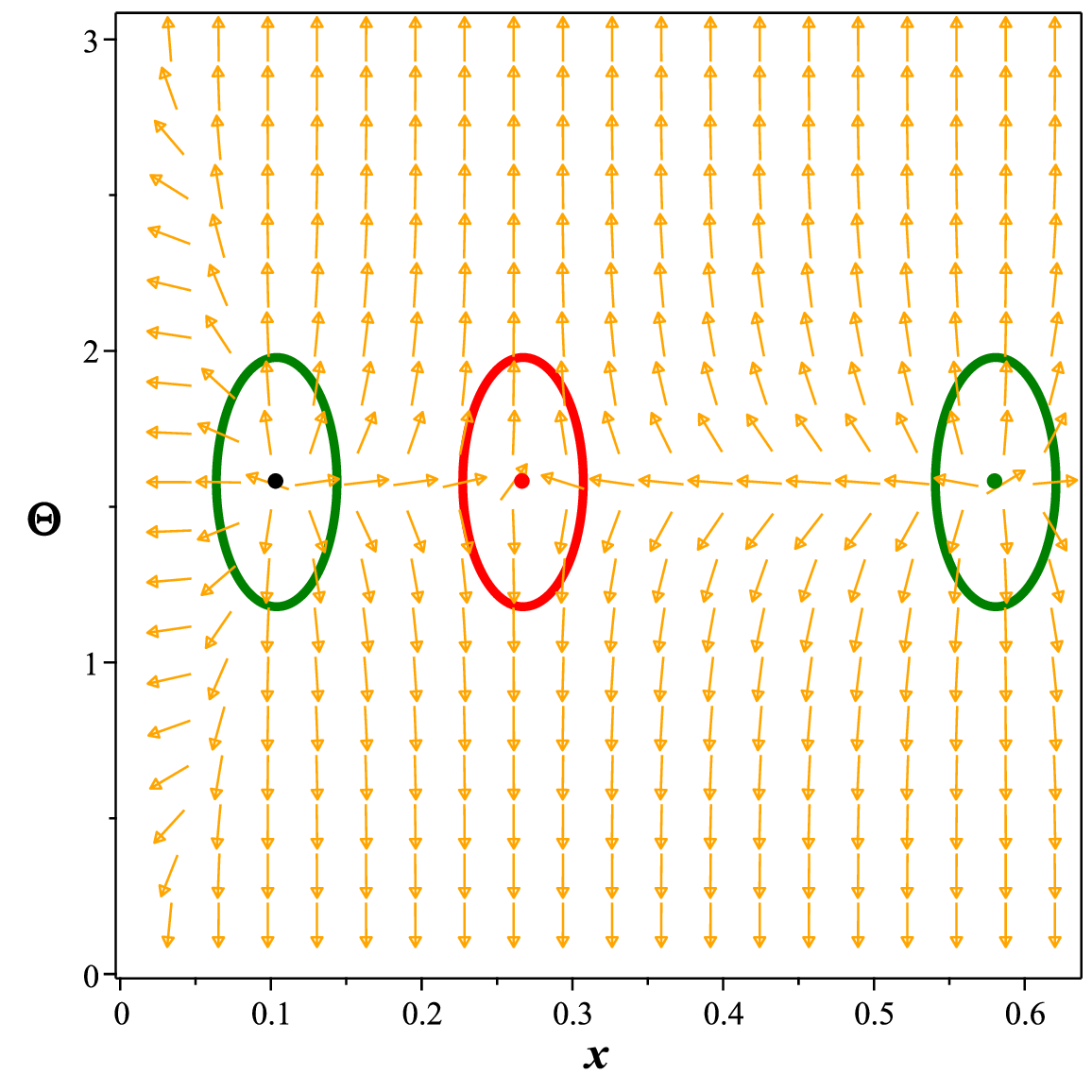}
 \label{300b}}
  \caption{\small{The curve in the ($x$–$\overline{ \tau}$) plot (left) and the corresponding zero points (ZPs) in the ($x$–$\Theta$) plot (right) are analyzed under specific parameter conditions.
We set $\overline{Q}=0.001$, and constants $k=Z_0=\gamma=\lambda_3=R=r_0=r_F=A=\omega_{k,d-1}=1$, with $z = \frac{3}{2}$, $d = 4$, $C = 7.9$, and $\theta = 0.5$.
The analysis is conducted at a fixed temperature parameter $\overline{ \tau} = 2.86$.
The ZPs are located at coordinates ($x$–$\Theta$): (0.1038127759, $\frac{\pi}{2}$), (0.2672276612, $\frac{\pi}{2}$), and (0.5807312771, $\frac{\pi}{2}$).
These results indicate the presence of a "first-order phase transition" in the topological structure of the system.
}}
 \label{ّm3}
\end{center}
 \end{figure}
\begin{figure}[H]
 \begin{center}
 \subfigure[]{
 \includegraphics[height=6cm,width=7cm]{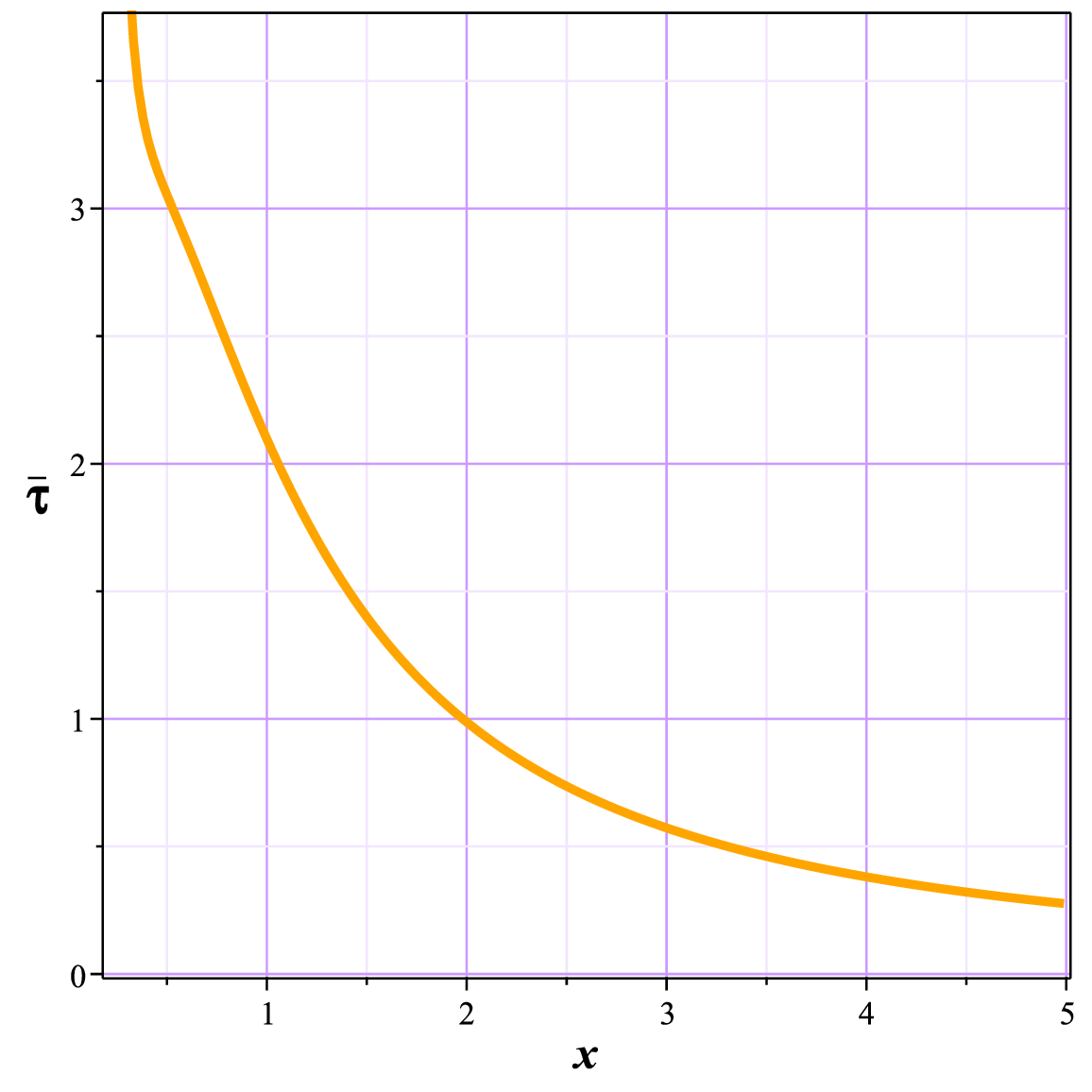}
 \label{400a}}
 \subfigure[]{
 \includegraphics[height=6cm,width=7cm]{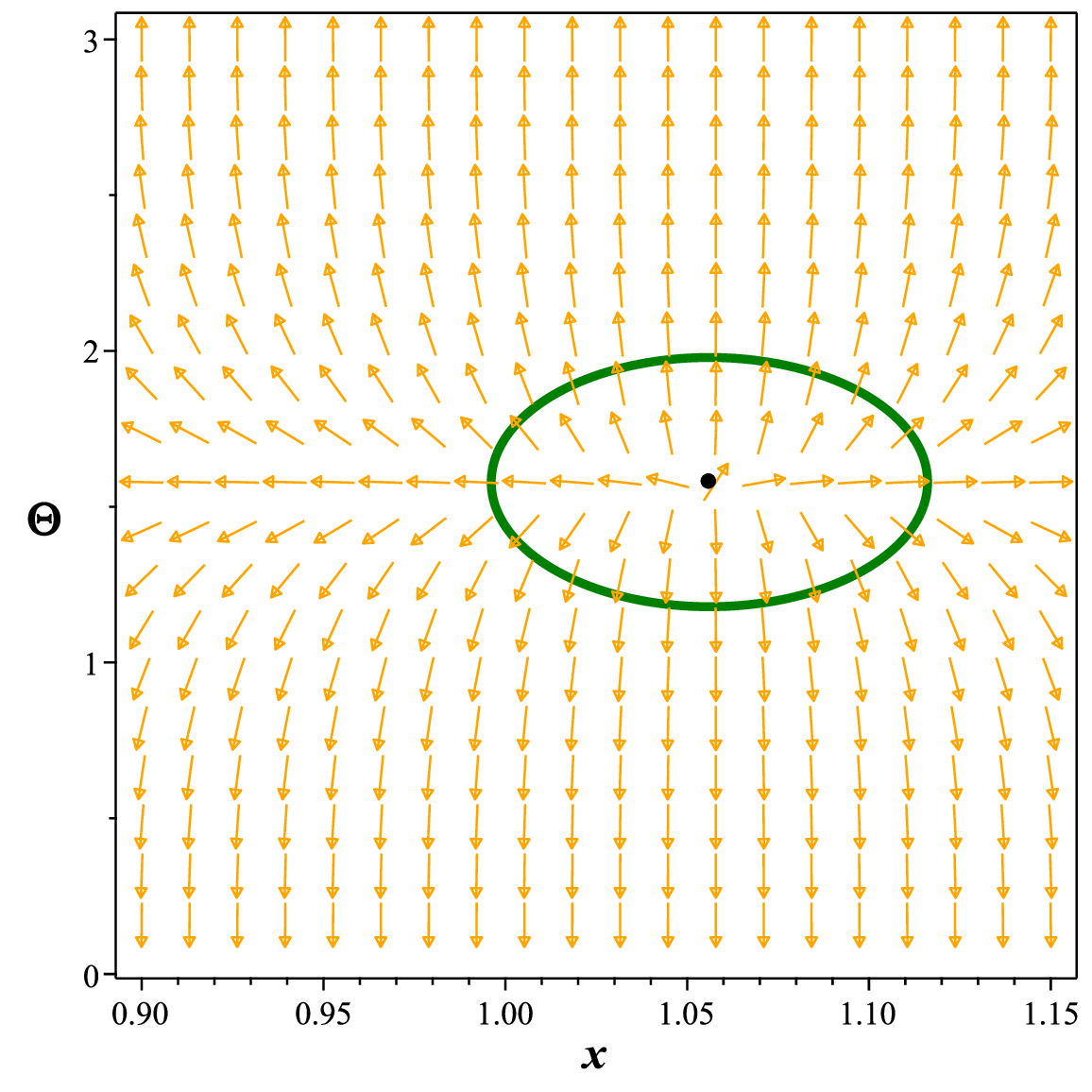}
 \label{400b}}
  \caption{\small{The curve in the ($x$–$\overline{ \tau}$) plot (left) and the corresponding zero point (ZP) in the ($x$–$\Theta$) plot (right) are analyzed under a specific parameter regime.
We use $\overline{Q} = 0.001$ with constants $k = Z_0 = \gamma = \lambda_3 = R = r_0 = r_F = A = \omega_{k,d-1} = 1$, and set $z = \frac{3}{2}$, $d = 4$, $C = 0.9$, $\theta = 0.5$.
The analysis focuses on the thermodynamic behavior at $\overline{ \tau} = 2$.
A single zero point is observed at the coordinate ($x$–$\Theta$) = (1.056127442, $\frac{\pi}{2}$).
This configuration indicates the presence of a "second-order phase transition" in the system.
}}
 \label{ّm4}
\end{center}
 \end{figure}
\subsection{$\theta=0$, $C_{crt}=6.613$ }
  \begin{figure}[H]
 \begin{center}
 \subfigure[]{
 \includegraphics[height=6cm,width=7cm]{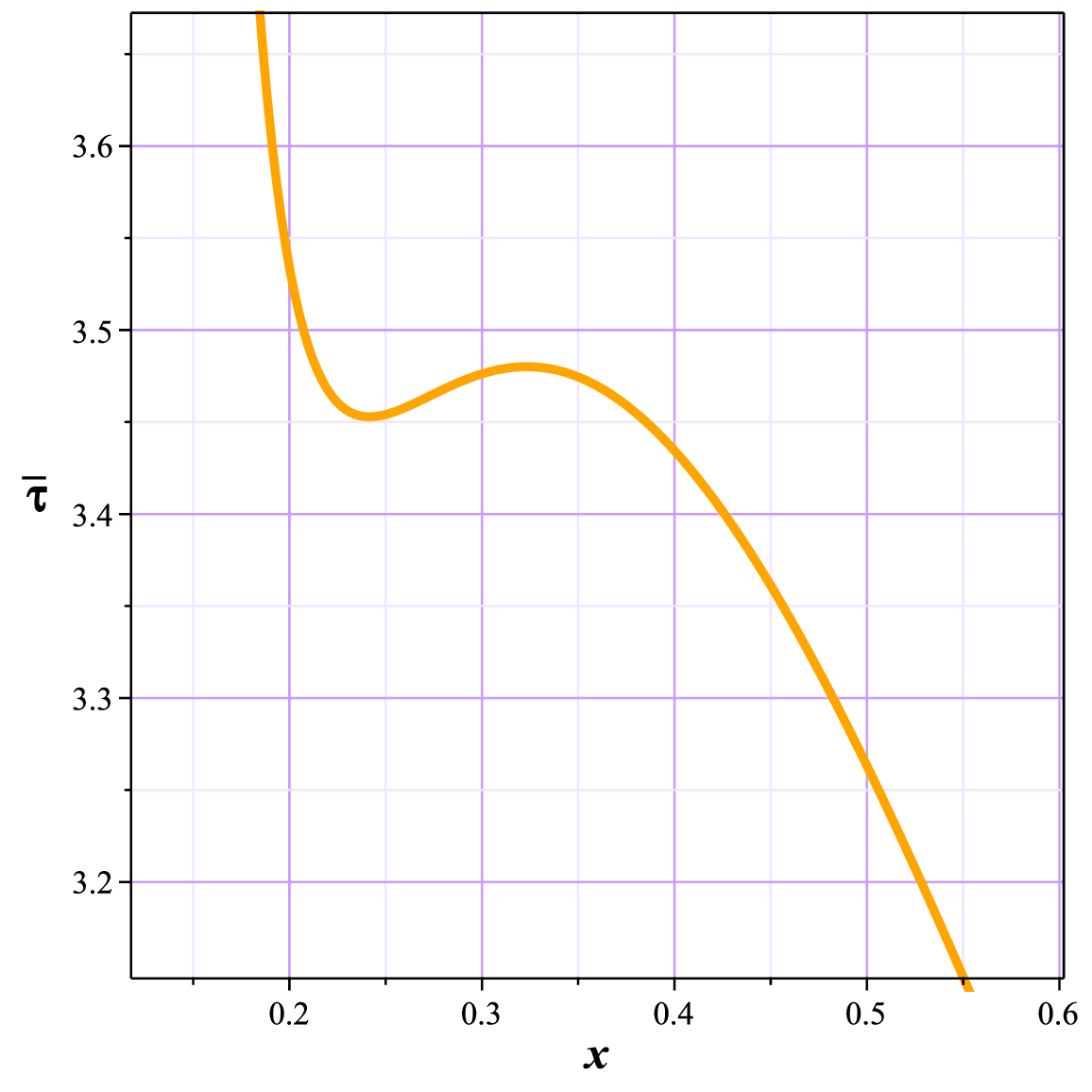}
 \label{500a}}
 \subfigure[]{
 \includegraphics[height=6cm,width=7cm]{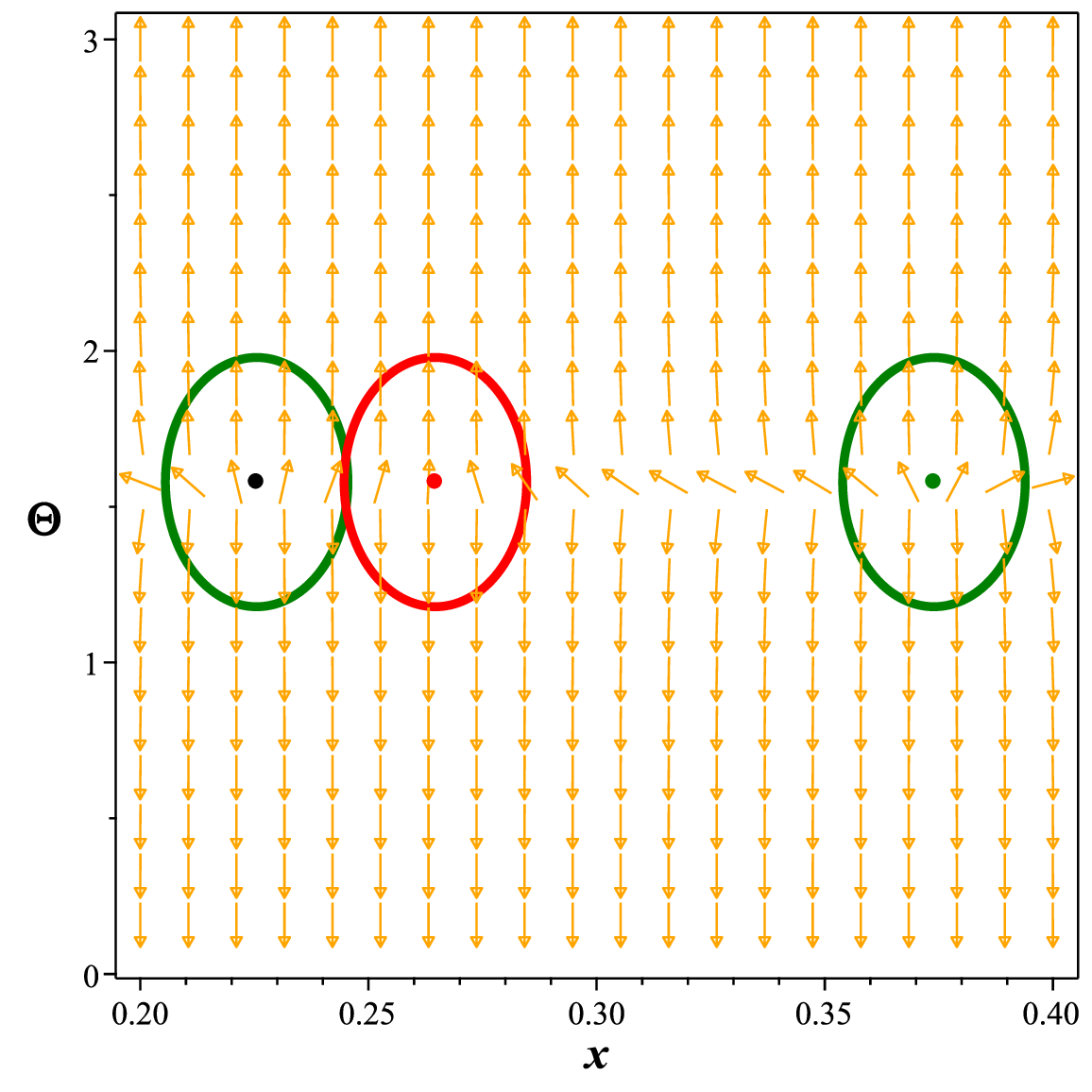}
 \label{500b}}
  \caption{\small{The behavior of the curve in the ($x$–$\overline{ \tau}$) plot (left) and the corresponding zero points (ZPs) in the ($x$–$\Theta$) plot (right) is examined under specific thermodynamic conditions.
The parameters are fixed as $\overline{Q} = 0.001$, with $k = Z_0 = \gamma = \lambda_3 = R = r_0 = r_F = A = \omega_{k,d-1} = 1$, and $z = \frac{3}{2}$, $d = 4$, $C = 7.9$, $\theta = 0$.
The analysis is conducted at the temperature parameter $\overline{ \tau} = 3.46$.
Three zero points are observed at ($x$–$\Theta$) coordinates: (0.2254889554, $\frac{\pi}{2}$), (0.2646634051, $\frac{\pi}{2}$), and (0.3739296163, $\frac{\pi}{2}$).
This multiplicity of ZPs signifies the occurrence of a "first-order phase transition" in the system's topological thermodynamics.}}
 \label{ّm5}
\end{center}
 \end{figure}
 \begin{figure}[H]
 \begin{center}
 \subfigure[]{
 \includegraphics[height=6cm,width=7cm]{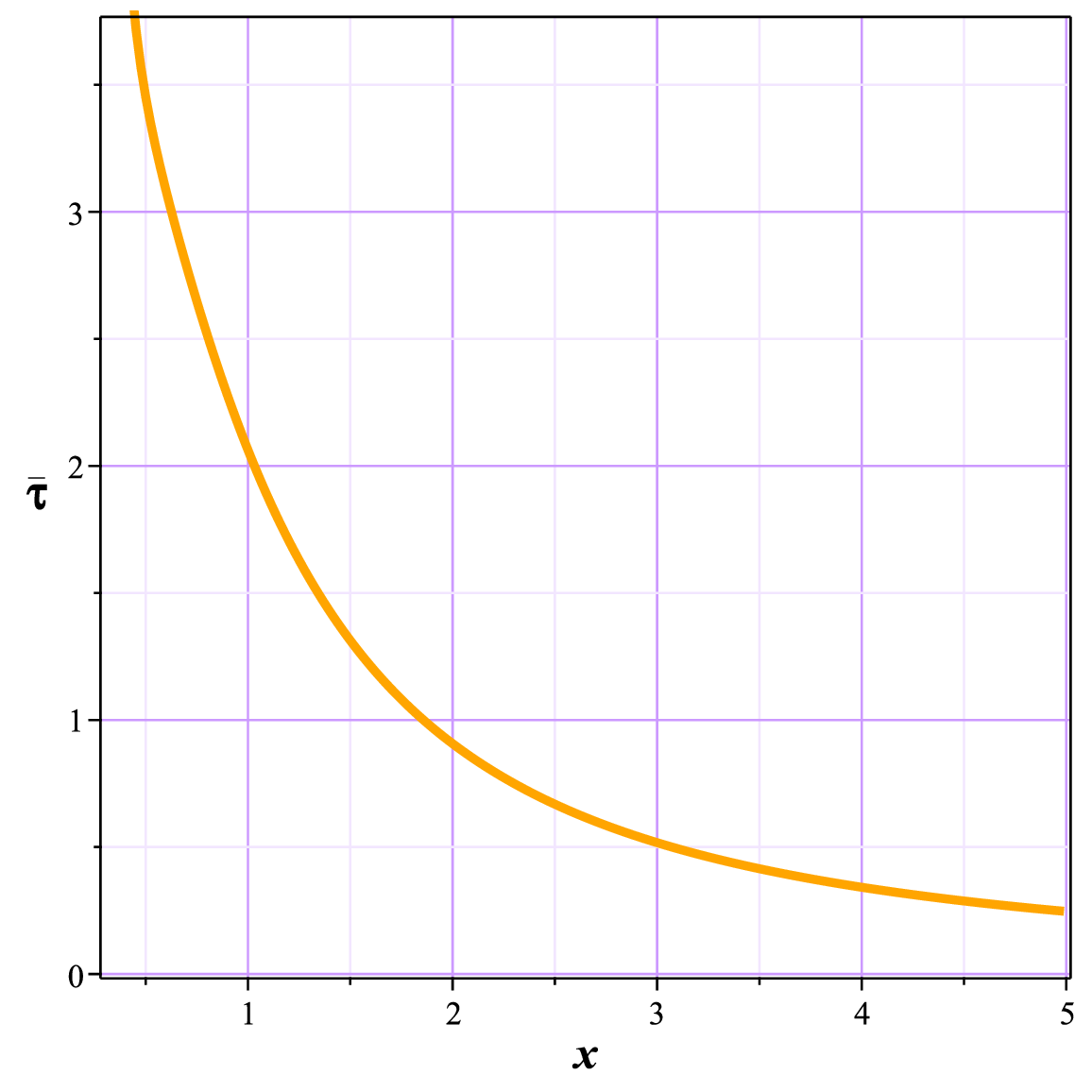}
 \label{600a}}
 \subfigure[]{
 \includegraphics[height=6cm,width=7cm]{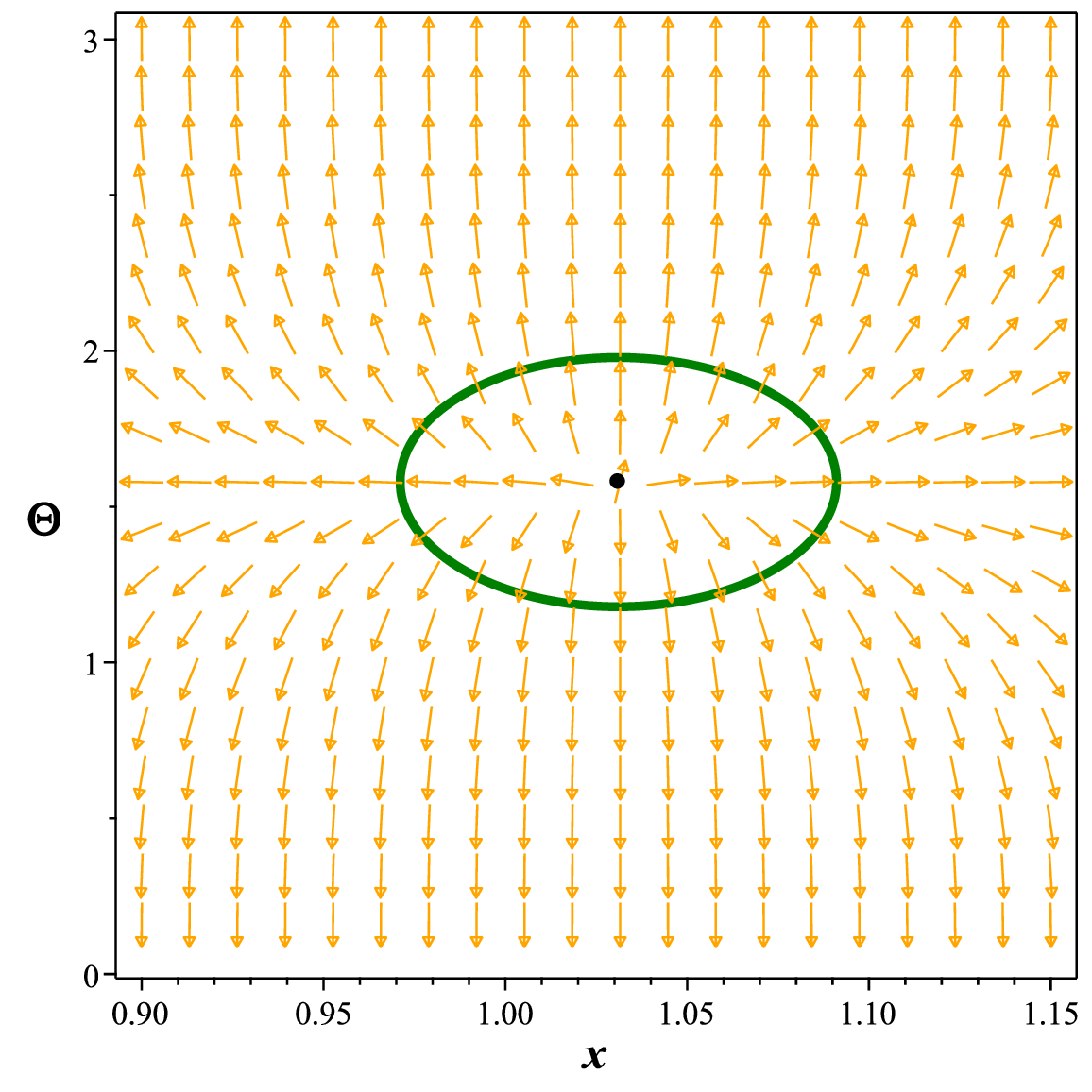}
 \label{600b}}
  \caption{\small{The behavior of the curve in the ($x$–$\overline{ \tau}$) plot (left) and the corresponding zero point (ZP) in the ($x$–$\Theta$) plot (right) is analyzed under controlled thermodynamic parameters.
We set $\overline{Q} = 0.001$, and take $k = Z_0 = \gamma = \lambda_3 = R = r_0 = r_F = A = \omega_{k,d-1} = 1$, with $z = \frac{3}{2}$, $d = 4$, $C = 0.9$, and $\theta = 0$.
The analysis is carried out at the fixed temperature value $\overline{ \tau} = 2$.
A single zero point is located at the coordinate ($x$–$\Theta$) = (1.031084358, $\frac{\pi}{2}$).
This isolated ZP indicates the system undergoes a "second-order phase transition", reflecting continuous but non-trivial topological change.}}
 \label{ّm6}
\end{center}
 \end{figure}
\subsection{$\theta=-0.5$, $C_{crt}=20.678$ }
  \begin{figure}[H]
 \begin{center}
 \subfigure[]{
 \includegraphics[height=6cm,width=7cm]{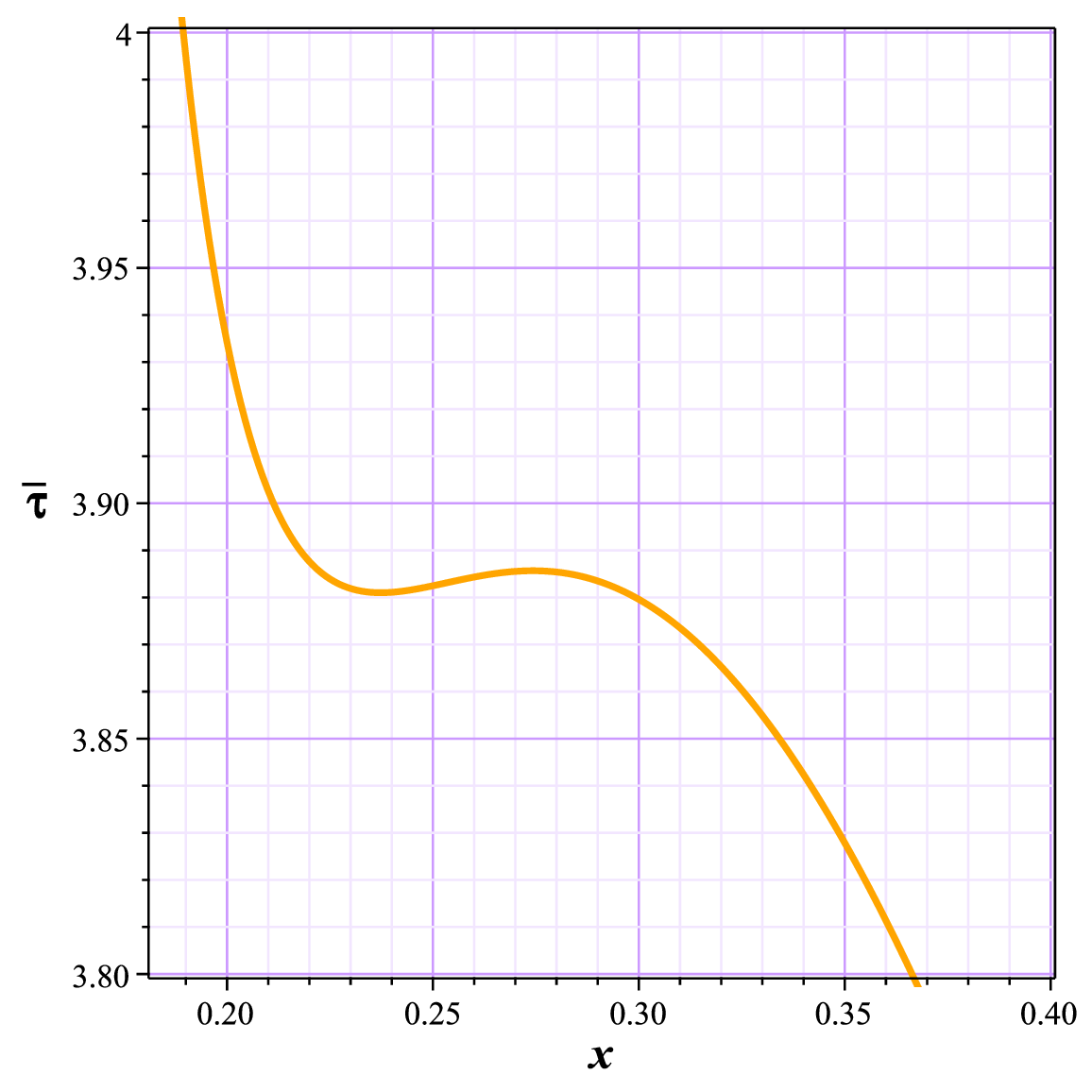}
 \label{700a}}
 \subfigure[]{
 \includegraphics[height=6cm,width=7cm]{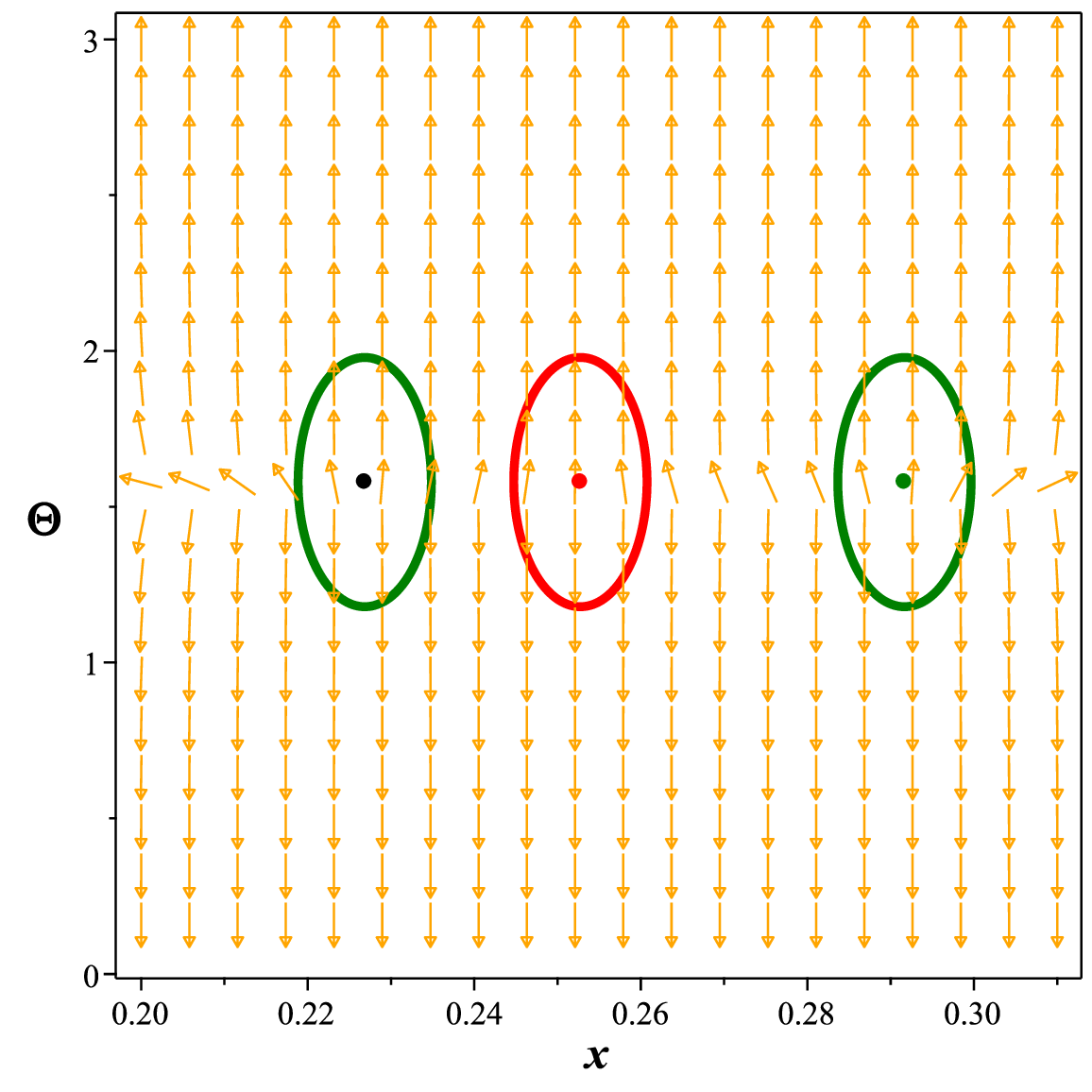}
 \label{700b}}
  \caption{\small{The evolution of the curve in the ($x$–$\overline{ \tau}$) plot (left) and the corresponding zero points (ZPs) in the ($x$–$\Theta$) plot (right) is studied under a specific thermodynamic setting.
The parameters are chosen as $\overline{Q} = 0.001$, with $k = Z_0 = \gamma = \lambda_3 = R = r_0 = r_F = A = \omega_{k,d-1} = 1$, and values $z = \frac{3}{2}$, $d = 4$, $C = 22$, and $\theta = -0.5$.
The system is evaluated at the temperature parameter $\overline{ \tau} = 2.3883$.
Three zero points are identified at ($x$–$\Theta$) coordinates: (0.2268480364, $\frac{\pi}{2}$), (0.2527415340, $\frac{\pi}{2}$), and (0.2916538929, $\frac{\pi}{2}$).
This presence of multiple ZPs clearly indicates a "first-order phase transition" in the topological behavior of the system.}}
 \label{ّm7}
\end{center}
 \end{figure}
 \begin{figure}[H]
 \begin{center}
 \subfigure[]{
 \includegraphics[height=6cm,width=7cm]{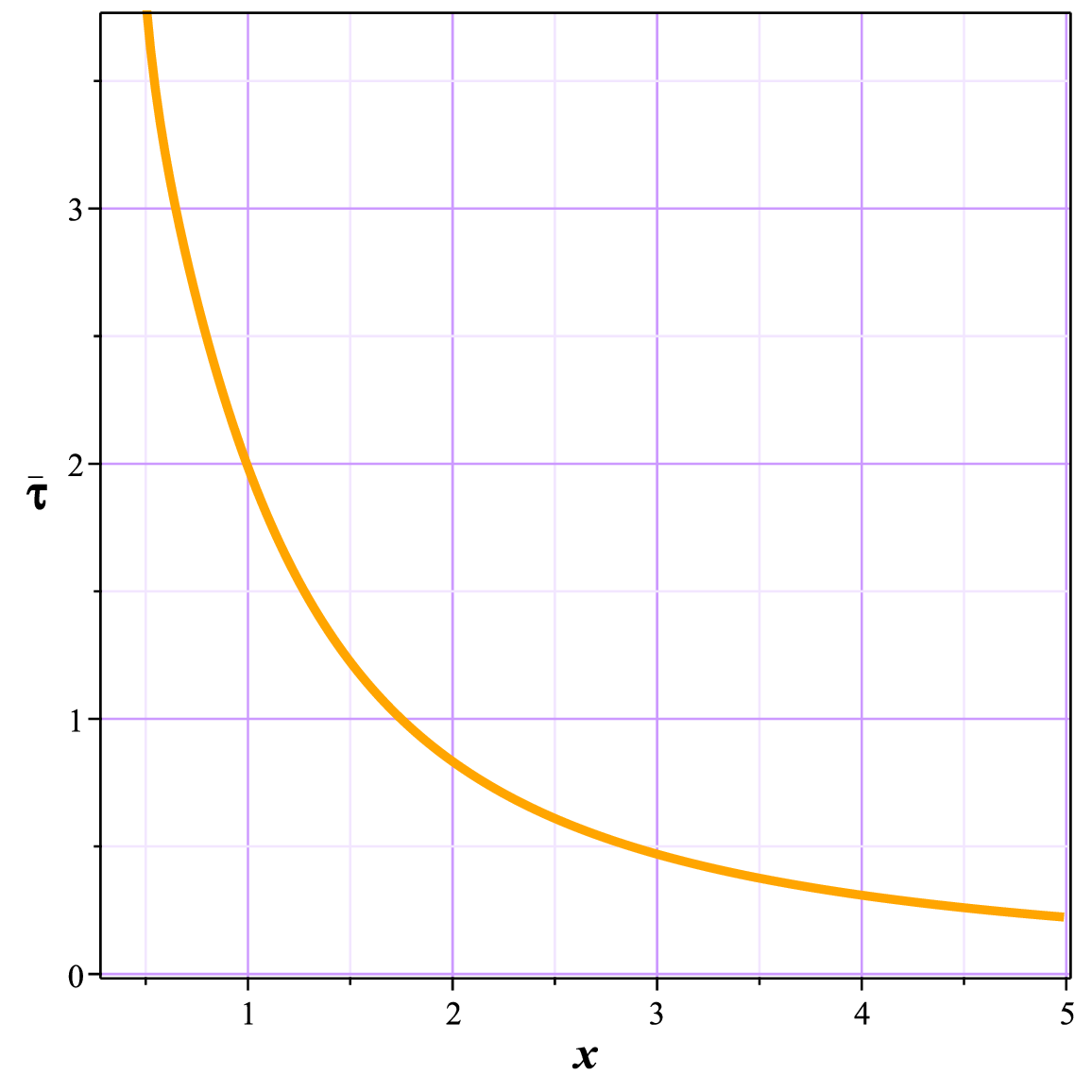}
 \label{800a}}
 \subfigure[]{
 \includegraphics[height=6cm,width=7cm]{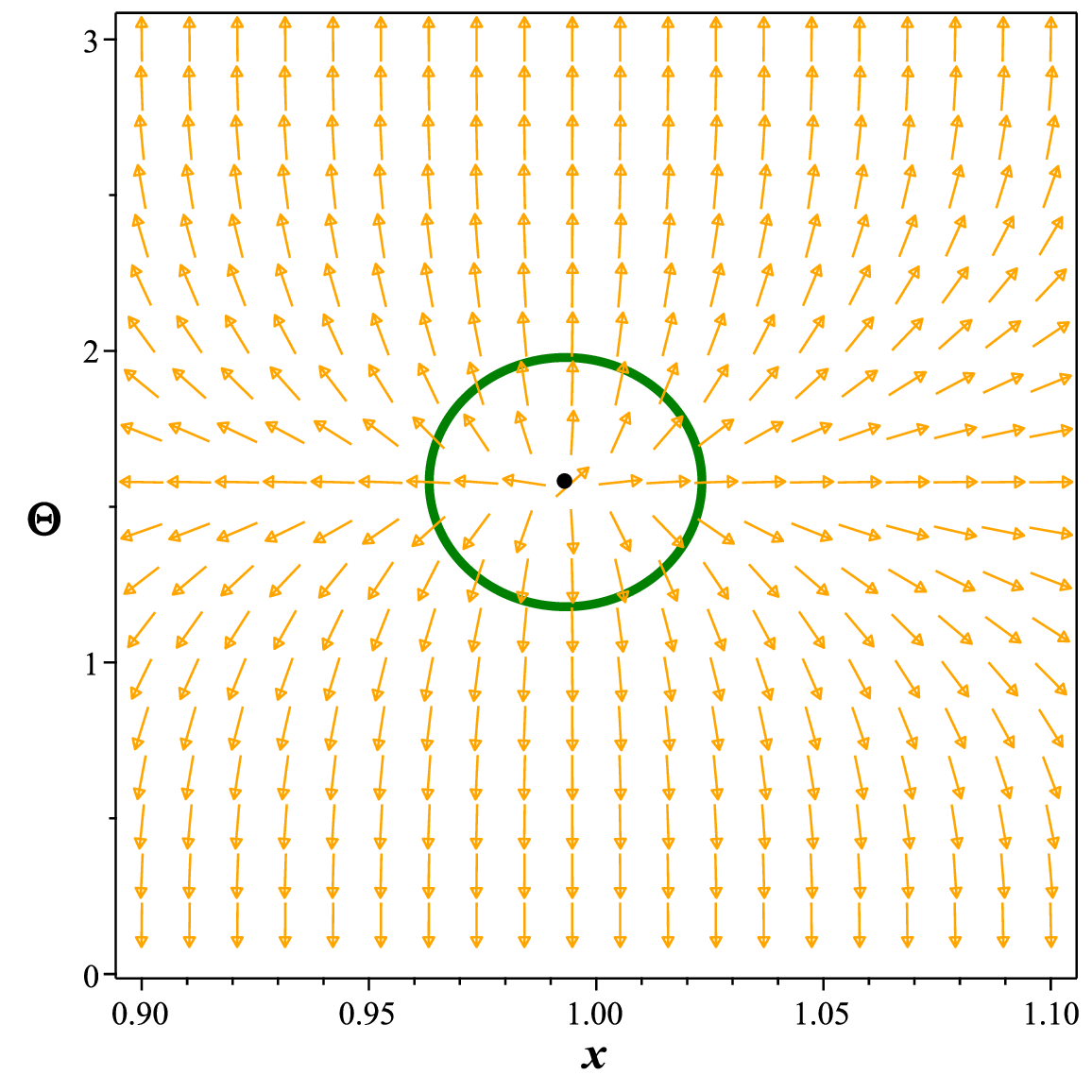}
 \label{800b}}
  \caption{\small{The behavior of the curve in the ($x$–$\overline{ \tau}$) plot (left) and the corresponding zero point (ZP) in the ($x$–$\Theta$) plot (right) is examined under a specific set of thermodynamic parameters.
The chosen parameters are $\overline{Q} = 0.001$, with $k = Z_0 = \gamma = \lambda_3 = R = r_0 = r_F = A = \omega_{k,d-1} = 1$, and fixed values $z = \frac{3}{2}$, $d = 4$, $C = 0.9$, and $\theta = -0.5$.
The analysis is performed at a temperature value of $\overline{ \tau} = 2$.
A single zero point is located at ($x$–$\Theta$) = (0.9932587205, $\frac{\pi}{2}$).
This solitary ZP signifies a second-order phase transition, indicating a continuous change in the system's topological structure.}}
 \label{ّm8}
\end{center}
 \end{figure}
As shown in Figs. 7 to 12, for different values of the free parameter—particularly $C$—the topological charges and classification of black holes are determined. In general, the system exhibits a total topological charge of \( W = +1 \) in all cases, with the variation occurring in the number of topological charges. This difference is reflected in the values of $C$. Specifically, when $C > C_{crt}$, the system possesses three topological charges (\(\omega = +1, -1, +1\)), with a total charge of \( W = +1 \), as seen in Figs. 7, 9, and 11 with a a first-order phase transition). Conversely, when $C < C_{crt}$, the number of topological charges is reduced, leaving only one topological charge (\(\omega = +1\)), as illustrated in Figs. 8, 10, and 12 with second-order phase transition), while the total charge remains \( W = +1 \). 
\section{Conclusion}
In this paper, our analysis will proceed through two distinct methodologies to provide a comprehensive understanding of their thermodynamic properties. First, we employed the classical approach, identifying critical points and examining the behavior of the free energy function as a function of temperature near the critical boundary. This method enables us to characterize phase transitions and assess the stability of these models, shedding light on their equilibrium states and the underlying mechanisms governing their thermal behavior. 

Next, to establish a comparative perspective and underscore the equivalence between these methodologies—particularly demonstrating the efficacy and accessibility of the topological approach—we extended our study to phase transitions through the formalism of topological charges. By leveraging this technique, we aim to refine our understanding of phase structures and thermodynamic properties in an intuitive and geometrically motivated manner.

Through this dual analysis, we provided a rigorous and insightful perspective on the thermodynamic attributes of Lifshitz black holes and HSV models, bridging classical and topological methodologies. Our findings will not only contribute to the broader discourse on black hole thermodynamics but also enhance the applicability of holographic techniques in exploring gravitational and quantum field theories. Through a systematic examination of the thermodynamic properties of black holes within the framework of conformal field theory (CFT), we have characterized the intricate phase transitions influenced by electric charge (\(\overline{Q}\)) and central charge (\(C\)). Our analysis, based on figures 1–4, revealed that for \(\overline{Q}<\overline{Q}_{crt}\), the free energy exhibits swallowtail behavior, indicating a first-order phase transition between two thermodynamically stable branches. As $\overline{Q}=\overline{Q}_{crt}$, the system undergoes a second-order phase transition, where a small stable black hole seamlessly transforms into a large stable black hole. Beyond this critical point (\(\overline{Q}>\overline{Q}_{crt}\)), no phase transition occurs. Similarly, examining variations in \(C\), we determined that for \(C>C_{crt}\), a first-order phase transition is observed, while at \(C=C_{crt}\), the system undergoes a second-order transition. When \(C<C_{crt}\), no transition takes place. Additionally, our stability analysis, as presented in Figures 5 and 6, highlighted an intriguing contrast in the structural behavior of \(\overline{Q}\) and \(C\). While the system exhibits localized instability for \(\overline{Q}<\overline{Q}_{crt}\) but remains stable otherwise, the opposite trend holds for \(C\), where stability is maintained except for a narrow region at \(C>C_{crt}\). Moving beyond classical thermodynamic descriptions, we explored the topological structure of black hole phase transitions via the computation of topological charges (Figures 7–12). We established that while the total topological charge (\(W = +1\)) remains conserved across all scenarios, the number of individual topological charges varies depending on \(C\). Specifically, when \(C>C_{crt}\), the system possesses three distinct topological charges (\(\omega = +1, -1, +1\)), corresponding to a first-order phase transition. Conversely, when \(C<C_{crt}\), the transition reduces to a second-order process with only a single topological charge (\(\omega = +1\)). A complementary classification based on \(\overline{Q}\) (Figures 13–18) reaffirms this structure: for \(\overline{Q}<\overline{Q}_{crt}\), the system exhibits three topological charges, indicating a first-order phase transition, whereas for \(\overline{Q}>\overline{Q}_{crt}\), only a single topological charge persists, marking a second-order transition. These findings collectively demonstrate a robust correspondence between classical thermodynamic analysis and topological characterization of black hole phase transitions. The consistency in observed transitions validates the applicability of topological charge methods
\section{Appendix}
As previously stated, due to the distinct roles of the parameters $\overline{Q}$ and $C$ in our analysis, we now provide explanations regarding the classification and number of topological charges associated with the black hole topology. As shown in Figs. 13 to 18, the topological charges and classification of black holes are determined based on different values of the free parameter—particularly $\overline{Q}$. In general, the system maintains a total topological charge of \( W = +1 \) in all cases, with the variation occurring in the number of individual topological charges. This difference is directly influenced by the values of $\overline{Q}$. Specifically, when $\overline{Q} < \overline{Q}_{crt}$, the system exhibits three topological charges (\(\omega = +1, -1, +1\)), while the total charge remains \( W = +1 \). This behavior is illustrated in Figs. 13, 15, and 17, which correspond to first-order phase transitions. Conversely, when $C < C_{crt}$, the number of topological charges is reduced, leaving only a single topological charge (\(\omega = +1\)), as shown in Figs. 14, 16, and 18, which are associated with second-order phase transitions, while the total charge remains \( W = +1 \).
\subsection{$\theta=0.5, \overline{Q}_{crt}=0.00269$}
 \begin{figure}[H]
 \begin{center}
 \subfigure[]{
 \includegraphics[height=5cm,width=7cm]{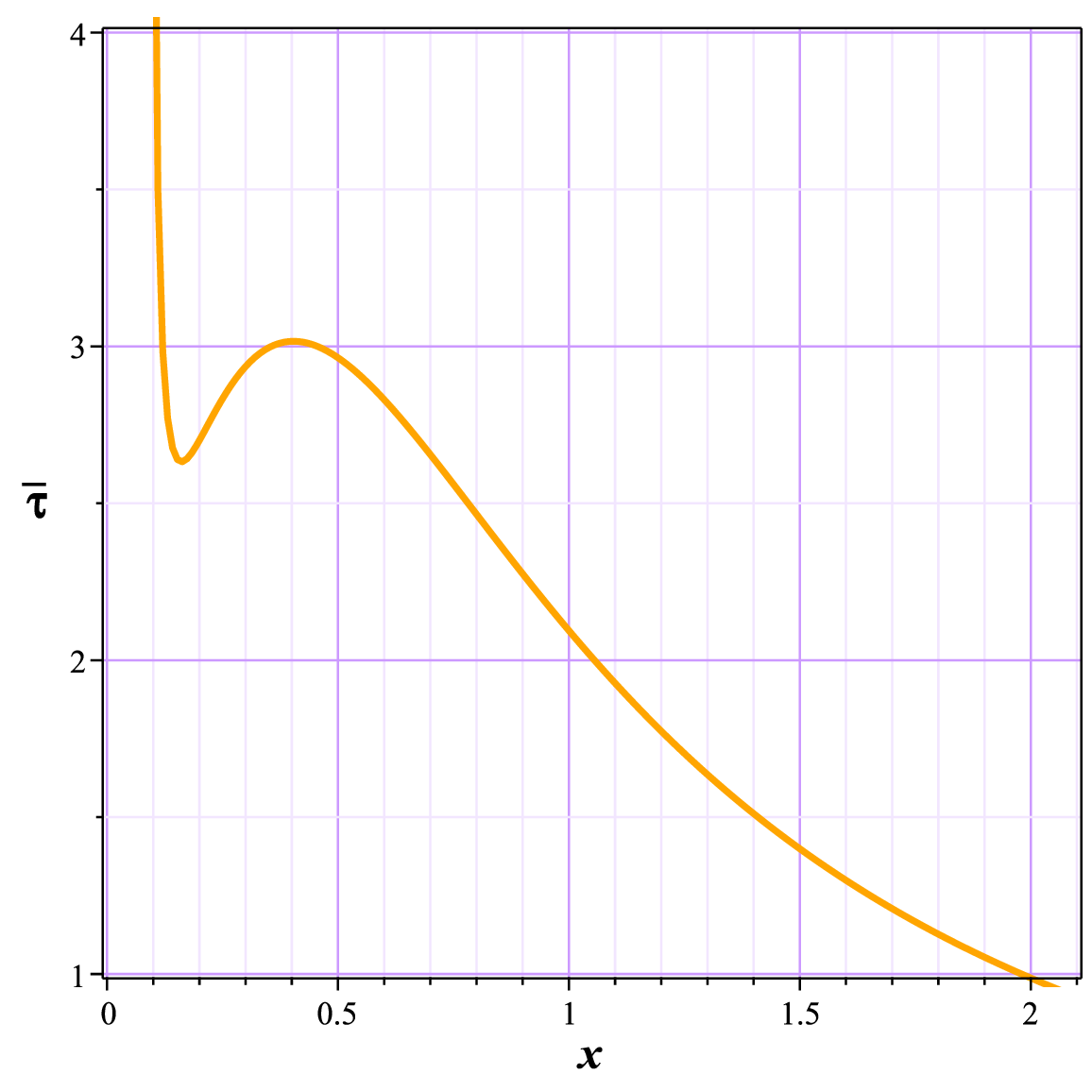}
 \label{910a}}
 \subfigure[]{
 \includegraphics[height=5cm,width=7cm]{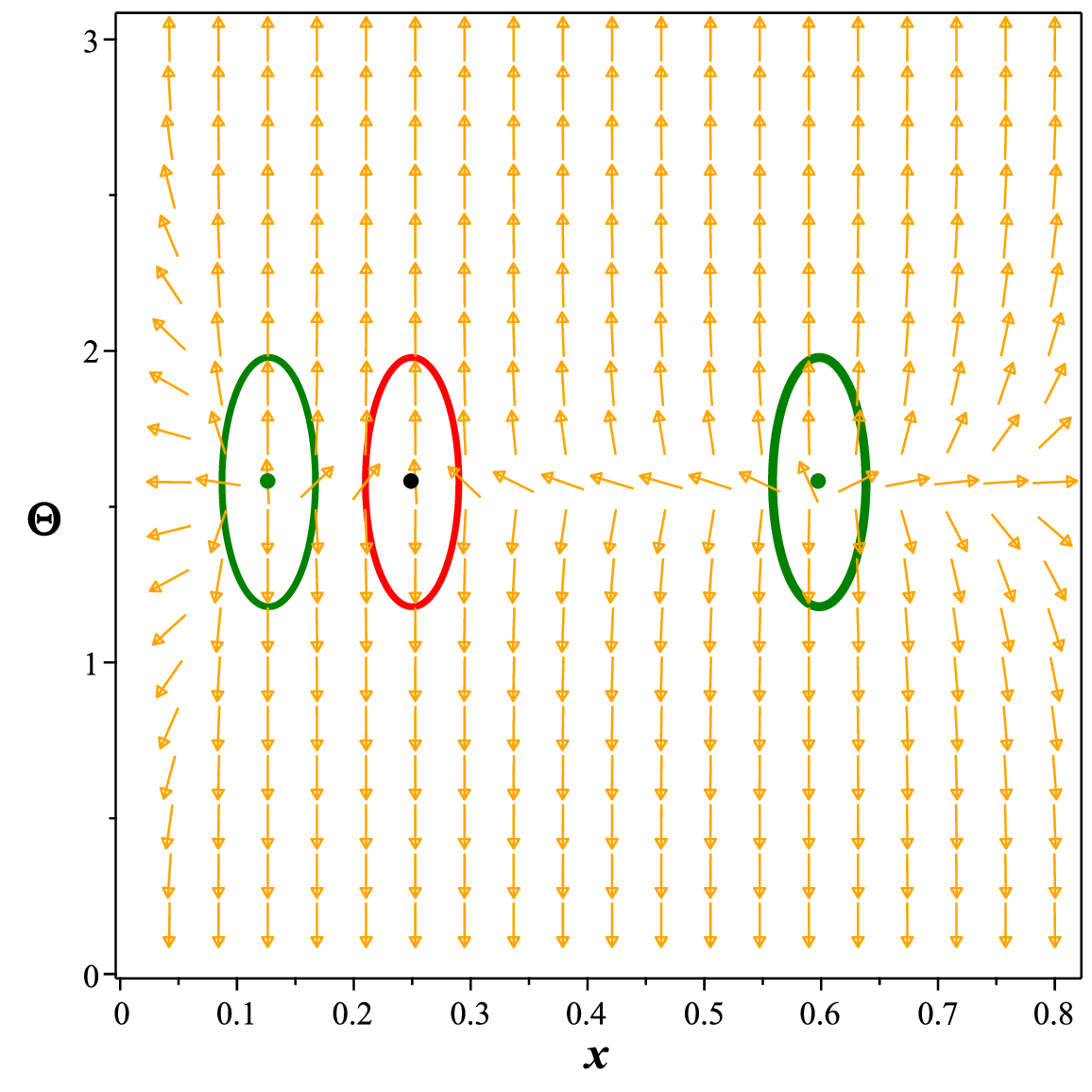}
 \label{910b}}
  \caption{\small{The curve in the ($x$–$\overline{ \tau}$) plot (left) and the zero points (ZPs) in the ($x$–$\Theta$) plot (right) are analyzed under specific parameter values.
Parameters are set as $C = 2$, with $k = Z_0 = \gamma = \lambda_3 = R = r_0 = r_F = A = \omega_{k,d-1} = 1$, along with $z = \frac{3}{2}$, $d = 4$, $\overline{Q} = 0.001$, and $\theta = 0.5$.
The system is studied at the temperature parameter $\overline{ \tau} = 2.833$.
Three zero points are located at coordinates ($x$–$\Theta$) = (0.1271852712, $\frac{\pi}{2}$), (0.2498976023, $\frac{\pi}{2}$), and (0.5983782517, $\frac{\pi}{2}$).
These multiple ZPs correspond to a "first-order phase transition", reflecting distinct topological changes in the system.}}
 \label{ّm10}
\end{center}
 \end{figure}
 
 \begin{figure}[H]
 \begin{center}
 \subfigure[]{
 \includegraphics[height=5cm,width=7cm]{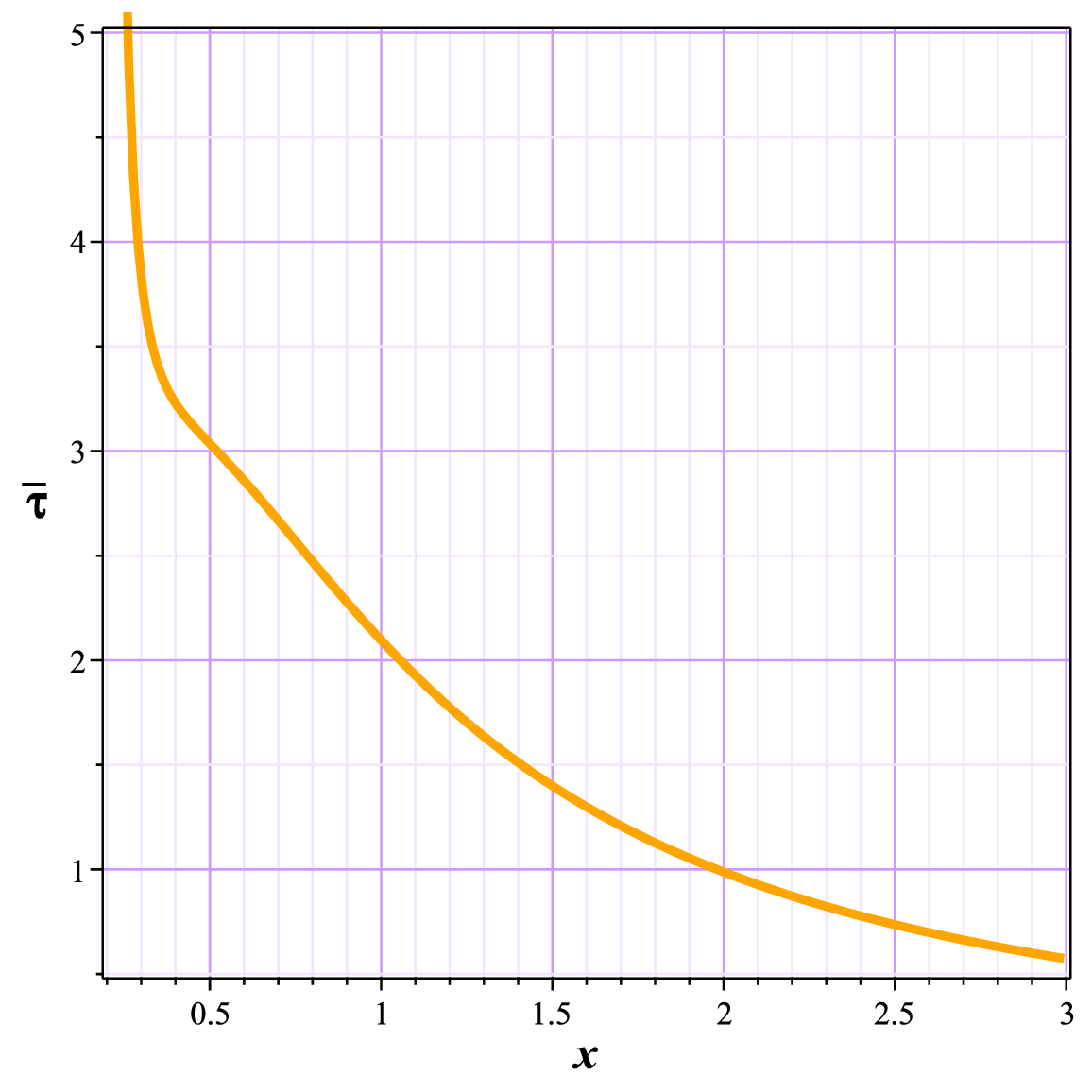}
 \label{900a}}
 \subfigure[]{
 \includegraphics[height=5cm,width=7cm]{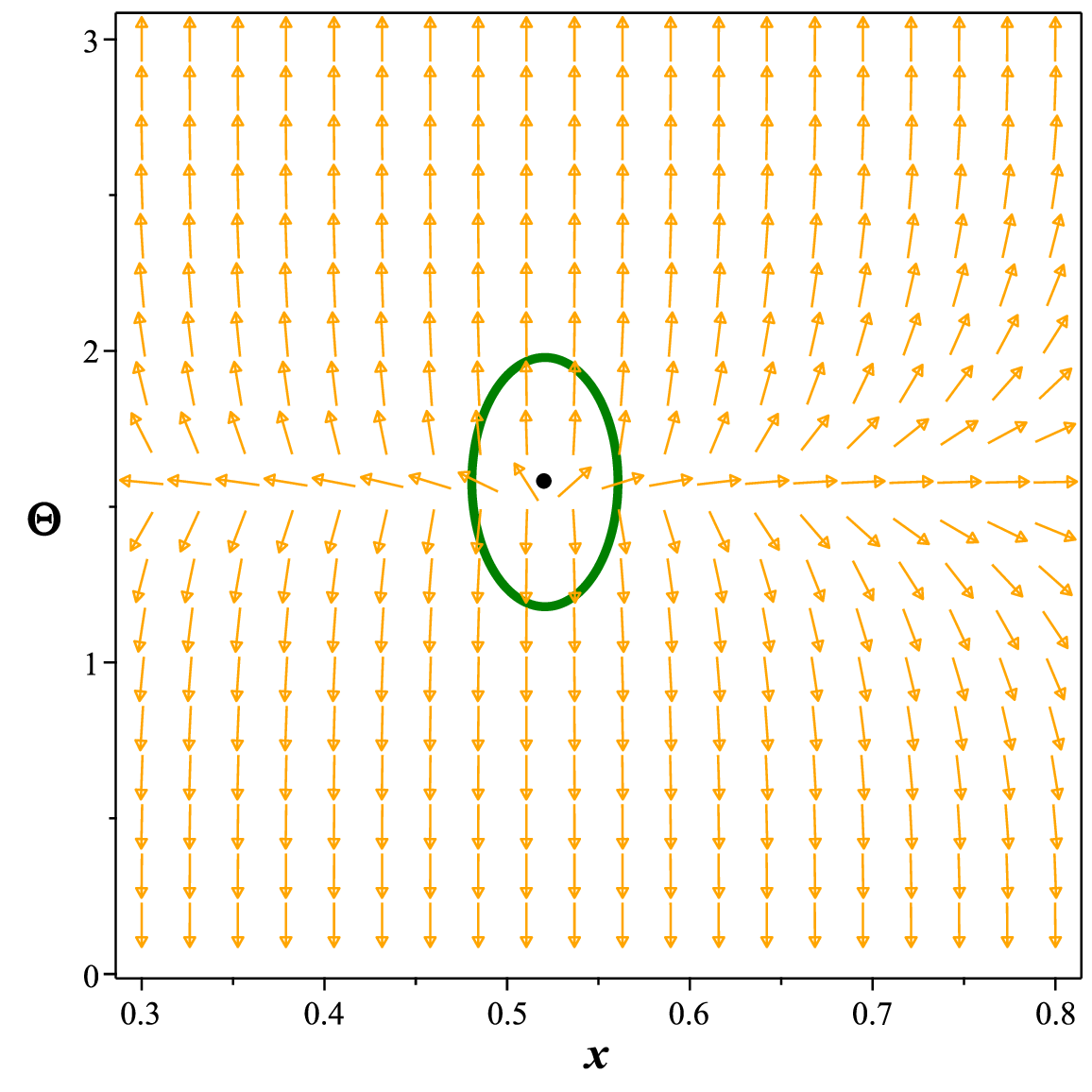}
 \label{900b}}
  \caption{\small{The curve in the ($x$–$\overline{ \tau}$) plot (left) and the zero point (ZP) in the ($x$–$\Theta$) plot (right) are analyzed under specific thermodynamic parameters.
We set $C = 2$, with constants $k = Z_0 = \gamma = \lambda_3 = R = r_0 = r_F = A = \omega_{k,d-1} = 1$, along with $z = \frac{3}{2}$, $d = 4$, $\overline{Q} = 0.006$, and $\theta = 0.5$.
The analysis is performed at the temperature parameter $\overline{ \tau} = 3$.
A single zero point is observed at the coordinate ($x$–$\Theta$) = (0.5206801068, $\frac{\pi}{2}$).
This solitary zero point indicates a "second-order phase transition", reflecting a continuous topological change in the system.}}
 \label{ّm9}
\end{center}
 \end{figure}
\subsection{$\theta=0, \overline{Q}_{crt}=0.000907$}
 \begin{figure}[H]
 \begin{center}
 \subfigure[]{
 \includegraphics[height=5cm,width=7cm]{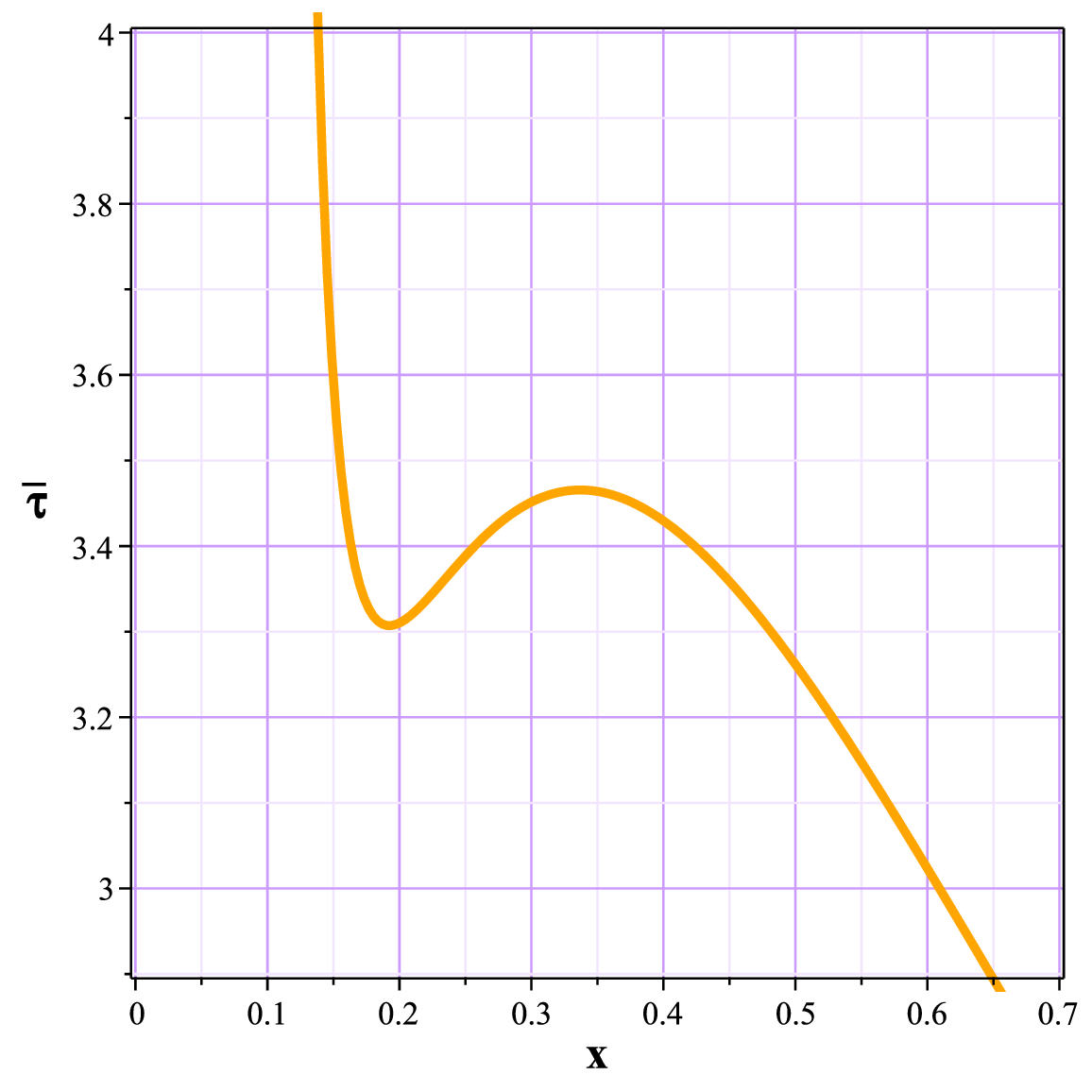}
 \label{920a}}
 \subfigure[]{
 \includegraphics[height=5cm,width=7cm]{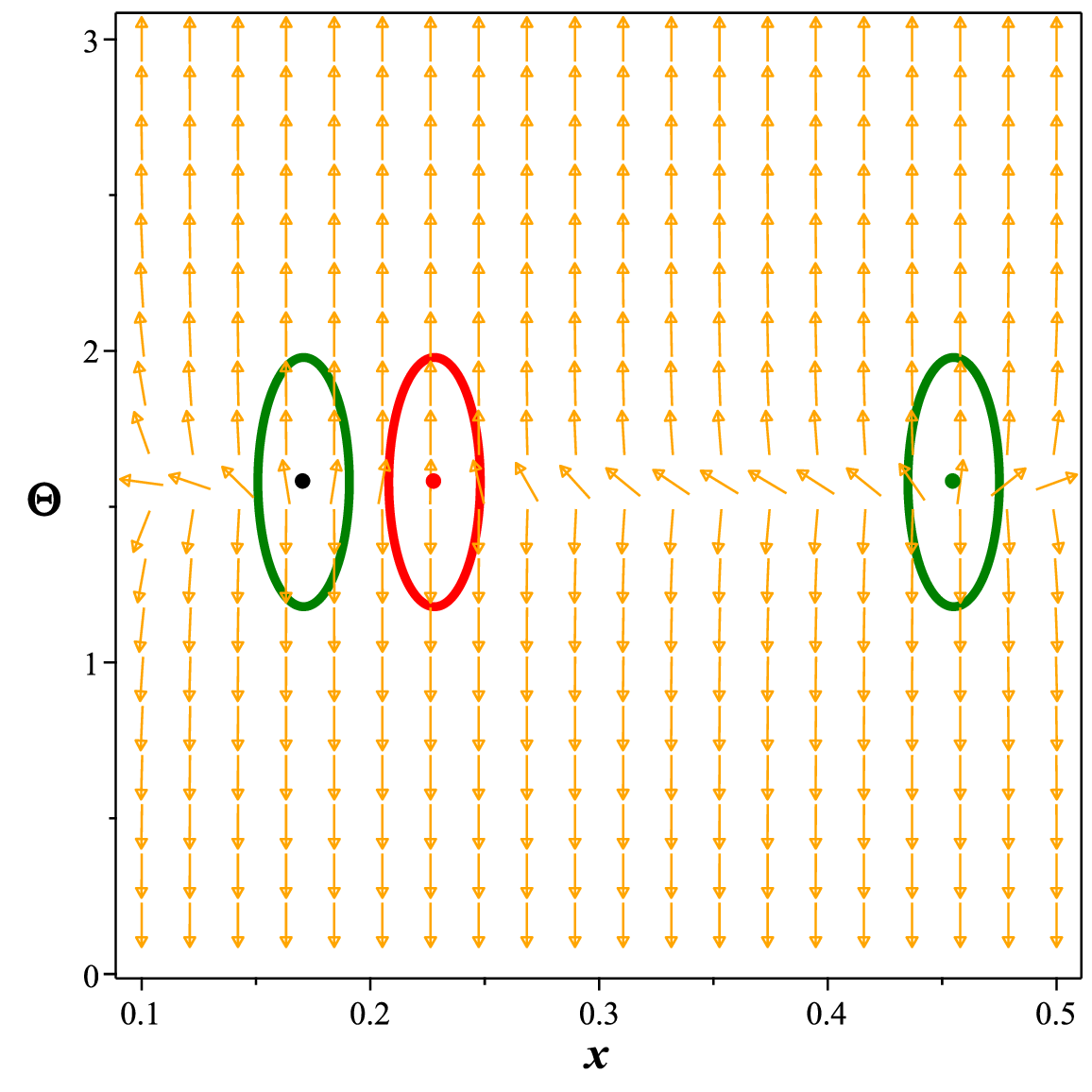}
 \label{920b}}
  \caption{\small{The curve in the ($x$–$\overline{ \tau}$) plot (left) and the zero points (ZPs) in the ($x$–$\Theta$) plot (right) are analyzed under the given parameter set.
Parameters are fixed as $C = 2$, with $k = Z_0 = \gamma = \lambda_3 = R = r_0 = r_F = A = \omega_{k,d-1} = 1$, and values $z = \frac{3}{2}$, $d = 4$, $\overline{Q} = 0.0005$, and $\theta = 0$.
The system is examined at temperature $\overline{ \tau} = 3.35$.
Three zero points appear at coordinates ($x$–$\Theta$) = (0.1708268874, $\frac{\pi}{2}$), (0.2280633172, $\frac{\pi}{2}$), and (0.4550223997, $\frac{\pi}{2}$).
The presence of these multiple zero points indicates a "first-order phase transition" in the topological structure of the black hole system}}
 \label{ّm11}
\end{center}
 \end{figure}
 
 \begin{figure}[H]
 \begin{center}
 \subfigure[]{
 \includegraphics[height=5cm,width=7cm]{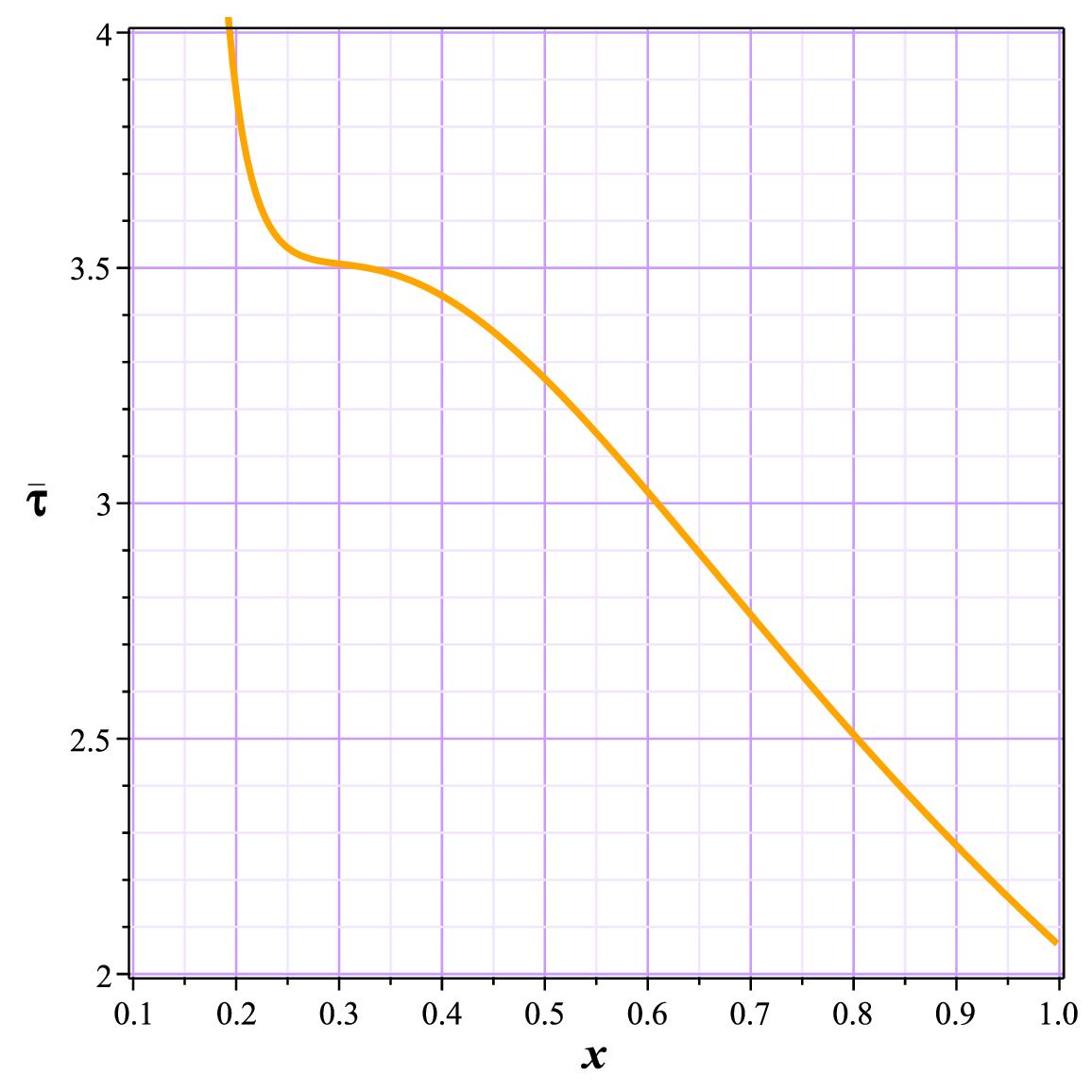}
 \label{930a}}
 \subfigure[]{
 \includegraphics[height=5cm,width=7cm]{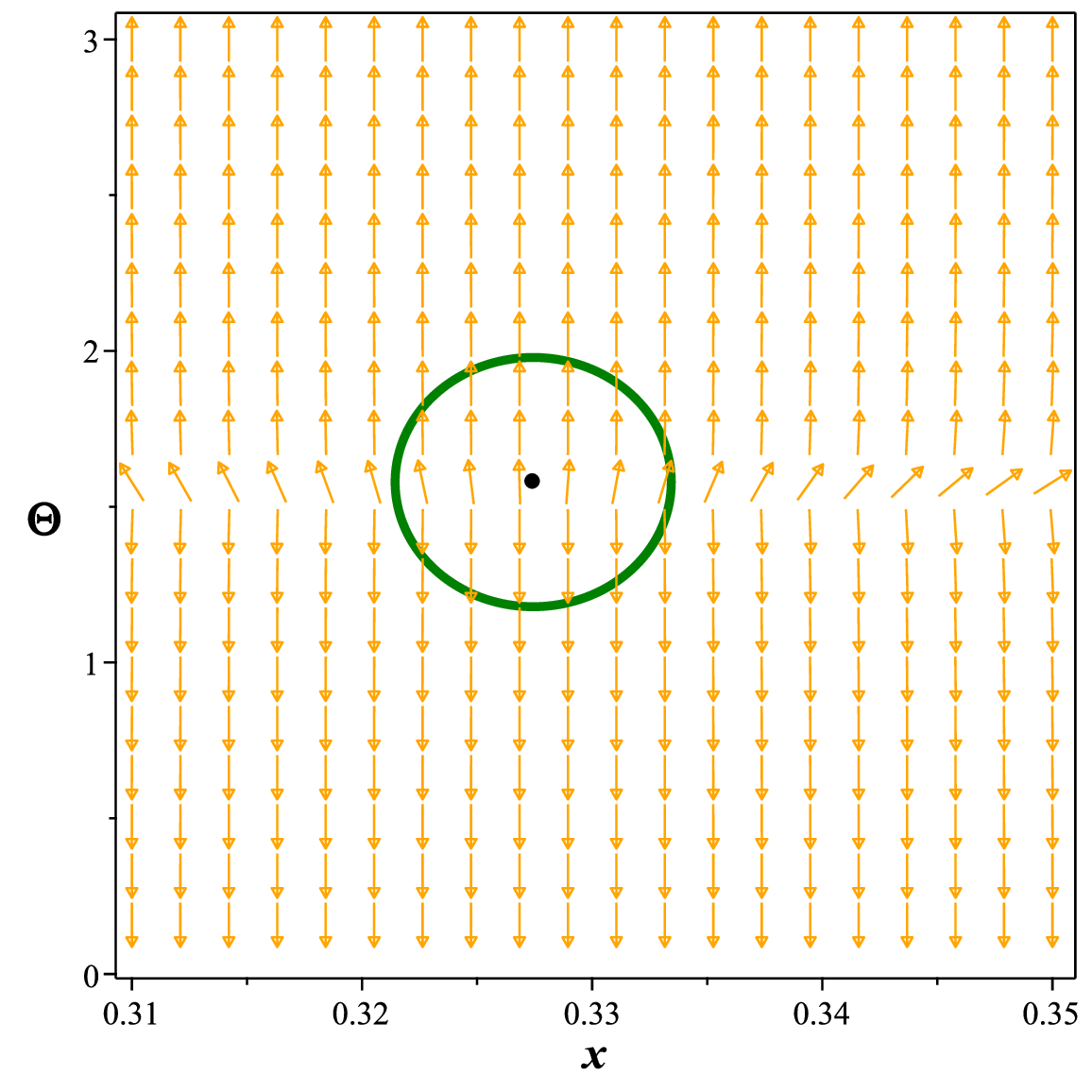}
 \label{930b}}
  \caption{\small{The curve in the ($x$–$\overline{ \tau}$) plot (left) and the zero point (ZP) in the ($x$–$\Theta$) plot (right) are studied under specific thermodynamic conditions.
Parameters are set as $C = 2$, with $k = Z_0 = \gamma = \lambda_3 = R = r_0 = r_F = A = \omega_{k,d-1} = 1$, and $z = \frac{3}{2}$, $d = 4$, $\overline{Q} = 0.001$, and $\theta = 0$.
The analysis is performed at the temperature $\overline{ \tau} = 3.5$.
A single zero point is identified at the coordinate ($x$–$\Theta$) = (0.3274380375, $\frac{\pi}{2}$).
This isolated zero point corresponds to a "second-order phase transition", indicating a continuous change in the system’s topological properties.}}
 \label{ّm12}
\end{center}
 \end{figure}

\subsection{$\theta=-0.5, \overline{Q}_{crt}=0.00029$}
 \begin{figure}[H]
 \begin{center}
 \subfigure[]{
 \includegraphics[height=5cm,width=7cm]{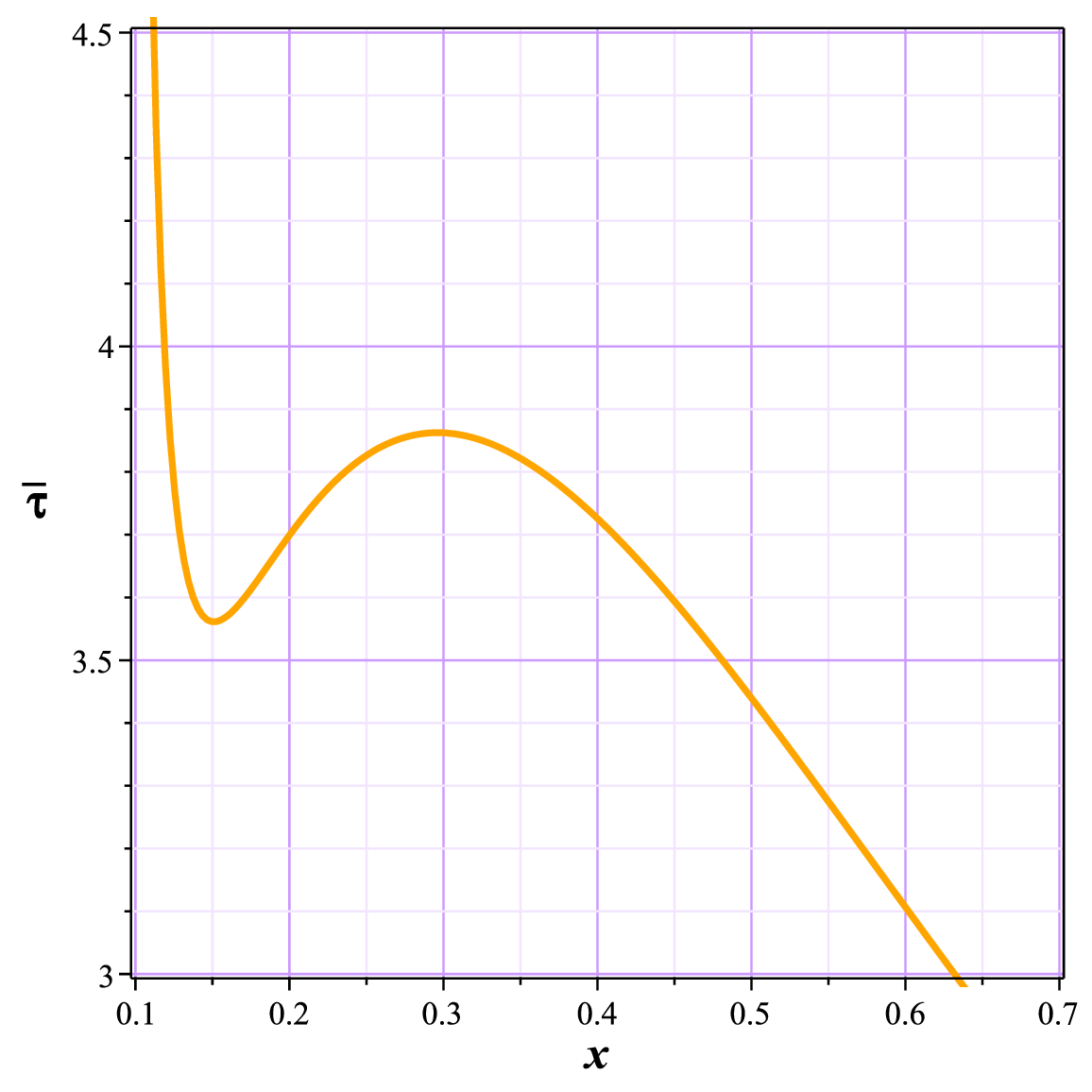}
 \label{940a}}
 \subfigure[]{
 \includegraphics[height=5cm,width=7cm]{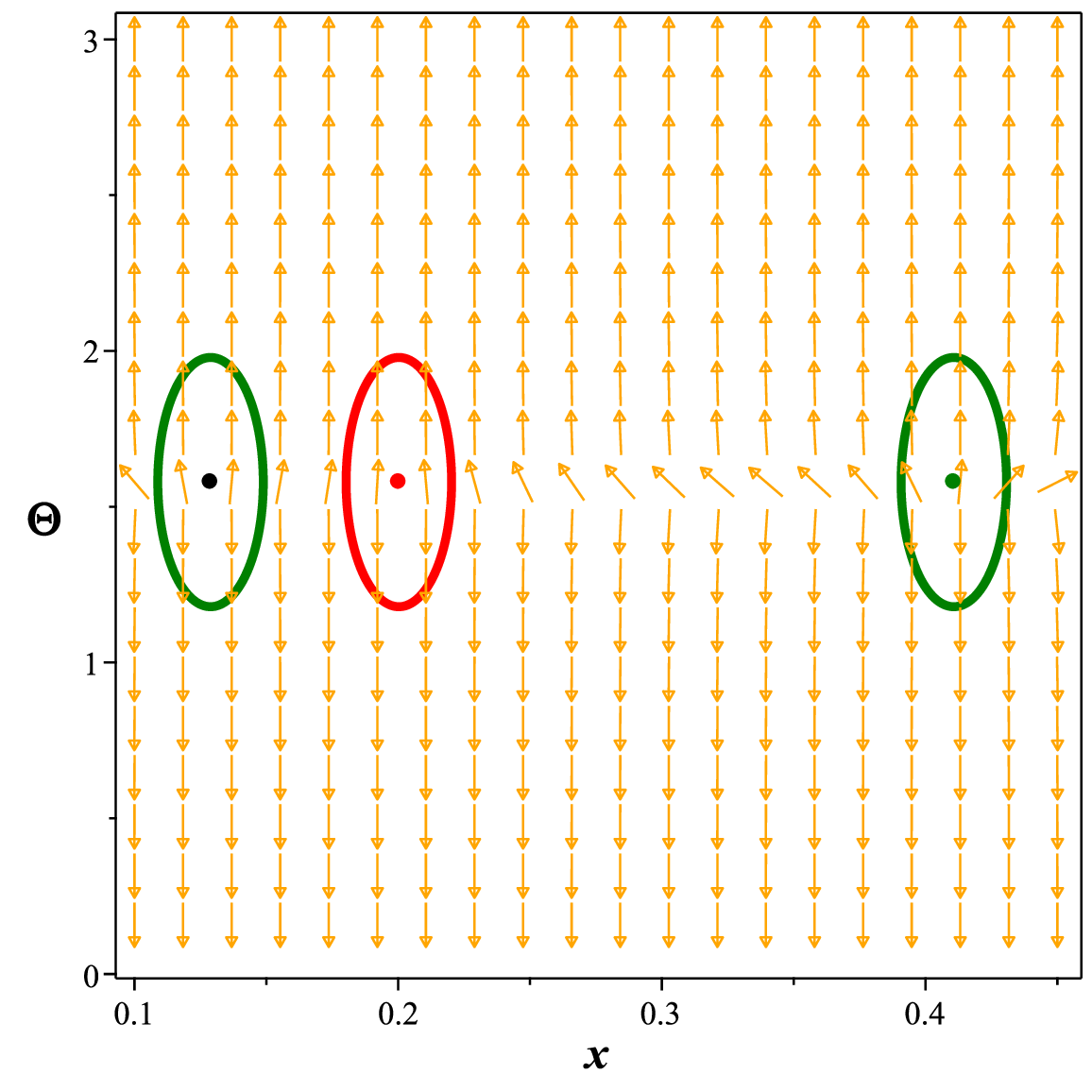}
 \label{940b}}
  \caption{\small{The curve in the ($x$–$\overline{ \tau}$) plot (left) and the zero points (ZPs) in the ($x$–$\Theta$) plot (right) are analyzed under specific parameter settings.
The parameters used are $C = 2$, with $k = Z_0 = \gamma = \lambda_3 = R = r_0 = r_F = A = \omega_{k,d-1} = 1$, and values $z = \frac{3}{2}$, $d = 4$, $\overline{Q} = 0.0001$, and $\theta = -0.5$.
The system is examined at temperature $\overline{ \tau}= 3.7$.
Three zero points are found at coordinates ($x$–$\Theta$) = (0.1288450470, $\frac{\pi}{2}$), (0.2002584712, $\frac{\pi}{2}$), and (0.4107284140, $\frac{\pi}{2}$).
The presence of multiple zero points indicates a "first-order phase transition" in the topological thermodynamics of the system.
}}
 \label{ّm13}
\end{center}
 \end{figure}
 
  \begin{figure}[H]
 \begin{center}
 \subfigure[]{
 \includegraphics[height=5cm,width=7cm]{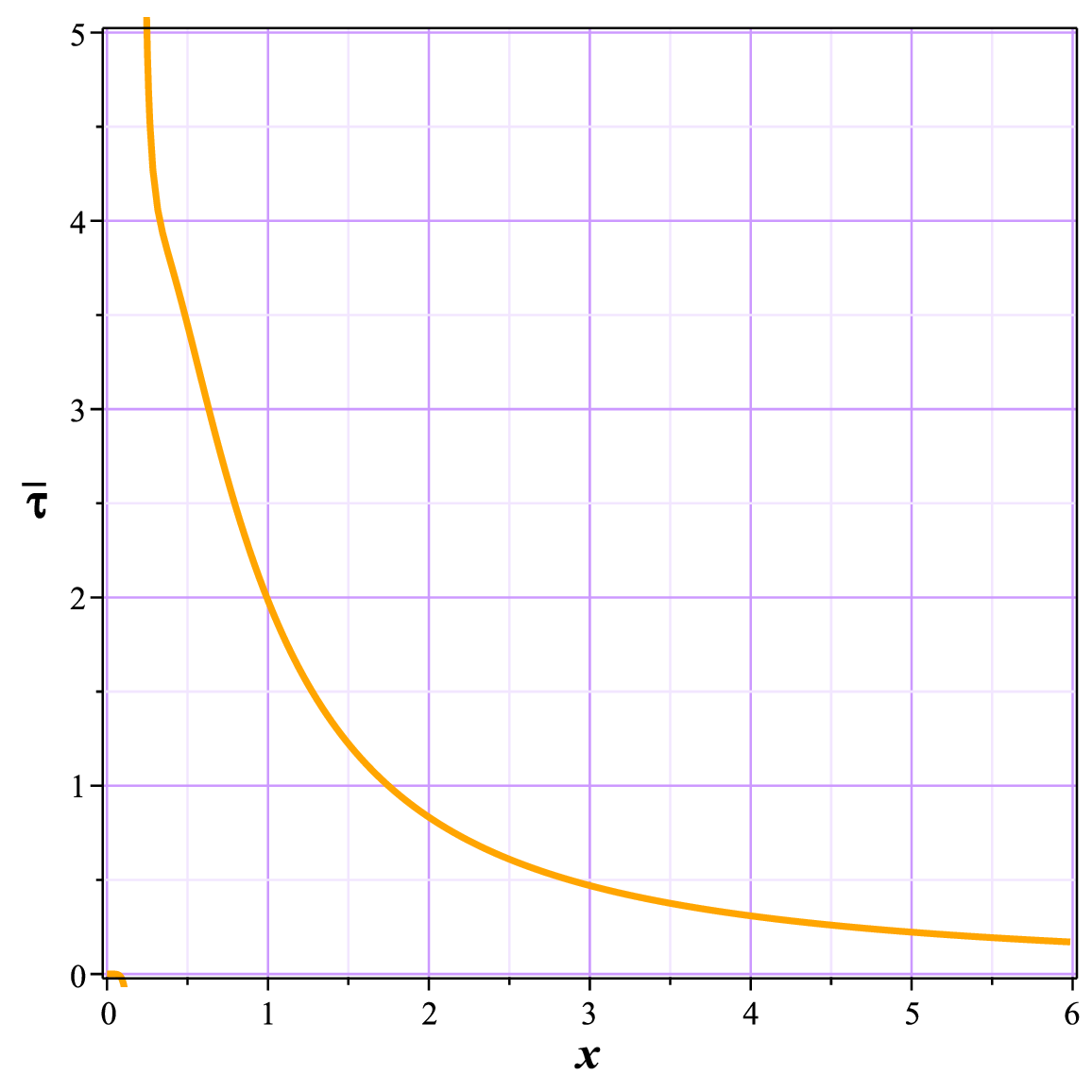}
 \label{950a}}
 \subfigure[]{
 \includegraphics[height=5cm,width=7cm]{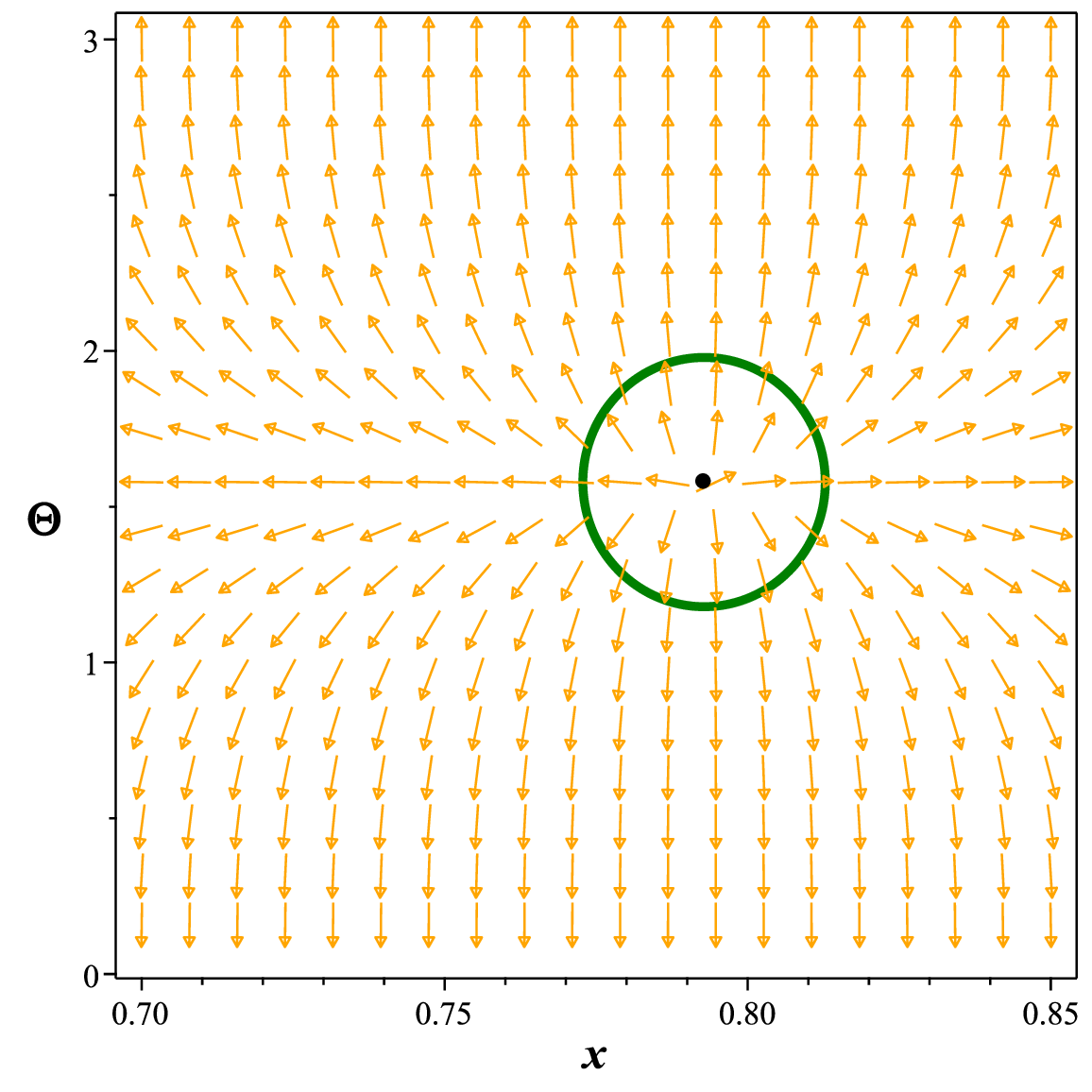}
 \label{950b}}
  \caption{\small{The curve in the ($x$–$\overline{ \tau}$) plot (left) and the zero point (ZP) in the ($x$–$\Theta$) plot (right) are examined under specific parameter values.
The parameters are set as $C = 2$, with $k = Z_0 = \gamma = \lambda_3 = R = r_0 = r_F = A = \omega_{k,d-1} = 1$, alongside $z = \frac{3}{2}$, $d = 4$, $\overline{Q} = 0.001$, and $\theta = -0.5$.
The analysis is conducted at the temperature $\overline{ \tau} = 2.5$.
A single zero point is located at ($x$–$\Theta$) = (0.7927951285, $\frac{\pi}{2}$).
This isolated zero point indicates a "second-order phase transition", representing a continuous topological change in the system.}}
 \label{ّm14}
\end{center}
 \end{figure}

\end{document}